\documentclass[prl,aps,reprint,footnotebib,amssymb,superscriptaddress]{revtex4-2}
\usepackage{amsmath}
\usepackage{bm}
\usepackage[dvipsnames]{xcolor}
\usepackage{graphicx}
\usepackage{nicematrix}
\usepackage{epstopdf}
\usepackage{epsfig}
\usepackage{amsfonts}
\usepackage[pdfencoding=auto,naturalnames]{hyperref}
\usepackage{hypcap}
\usepackage{mathtools}
\usepackage{verbatim}
\usepackage{tabularx}
\usepackage{bbm}
\usepackage{esvect}
\usepackage{subfigure}
\usepackage{slashed}
\usepackage{physics}
\usepackage{times}
\usepackage{multirow}
\usepackage{float}
\usepackage{cancel}

\hypersetup{colorlinks=true, citecolor=Blue, urlcolor=Blue, linkcolor=Blue, breaklinks=true}

\newcommand{\ssll}{\hspace{-1em}---}

\def\bea{\begin{eqnarray}}
\def\eea{\end{eqnarray}}
\def\be{\begin{equation}}
\def\ee{\end{equation}}

\def\ba{\begin{array}}
\def\ea{\end{array}}
\def\Tr{\mathrm{Tr}}

\begin{document}

\title{Boundary Criticality at the Nishimori Multicritical Point}

\author{Sheng Yang}
\thanks{S. Y. and X. S. contributed equally to this work.}
\affiliation{Institute for Advanced Study in Physics and School of Physics, Zhejiang University, Hangzhou 310058, China}

\author{Xinyu Sun}
\thanks{S. Y. and X. S. contributed equally to this work.}
\affiliation{Institute for Advanced Study, Tsinghua University, Beijing 100084, China}

\author{Shao-Kai Jian}
\email{sjian@tulane.edu}
\affiliation{Department of Physics and Engineering Physics, Tulane University, New Orleans, Louisiana 70118, USA}

\date{\today}

\begin{abstract}
We study boundary criticality at the Nishimori multicritical point of the two-dimensional (2D) random-bond Ising model. 
Using tensor-network methods, we construct a family of microscopic boundary conditions that incorporates both boundary-spin rotation and boundary disorder.
We identify three conformal boundary fixed points, corresponding to free, fixed, and random boundary conditions, and map out the boundary renormalization group flows among them. 
We extract the corresponding boundary conformal data, including the boundary entropies and the scaling dimensions of boundary primary operators, which characterize the boundary universality class. 
At the free boundary fixed point, we uncover the multifractal scaling of boundary spin fields. 
We further complement the numerical results with a controlled renormalization group analysis. 
Finally, we connect the boundary conformal data to quantum error-correcting codes, establishing a bridge between boundary universality class and boundary decoding threshold.
\end{abstract}

\maketitle

\paragraph{Introduction.}\ssll
Boundary conformal field theory (BCFT) provides a universal framework for boundary critical phenomena, classifying conformal boundary conditions through boundary states, boundary operators, and fusion rules~\cite{cardy1984conformal,cardy1989boundary,cardy1991bulk}. 
A key diagnostic is the Affleck--Ludwig boundary entropy~\cite{affleck1991universal}, which decreases along renormalization group (RG) flows in unitary theories~\cite{friedan2004boundary,casini2016the}, but need not be monotonic in disordered or nonunitary systems~\cite{Patrick2000g,RunkelThesis2000,Ashida2024System}. 
Such systems can host new boundary universality classes and multifractal boundary observables, as in random Potts models, localization transitions, and quantum Hall critical points~\cite{Christophe2000Universality,palagyi2000boundary,subramaniam2006surface,Mildenberger2007Boundary,Babkin2023Generalized,subramaniam2008boundary,obuse2008boundary}. 

The random-bond Ising model (RBIM) on the Nishimori line is a paradigmatic example where quenched disorder and enlarged symmetries define a distinguished critical manifold~\cite{nishimori1980exact,nishimori1981internal,Georges1985ExactII,pierre1988location}. 
Its multicritical point is believed to be described by a nonunitary CFT with enhanced supersymmetry~\cite{Georges1985ExactI,gruzberg2001random}, and its bulk critical behavior has been extensively studied~\cite{ozeki1987phase,singh1991spin,singh1996high,aarao1999universality,honecker2001universality,merz2002two,Queiroz2003Correlation,hasenbusch2008,Queiroz2009Location,wang2014topologically,chen2025tensor,delfino2025critical,delfino2025exact,agrawal2024dynamics}. 
By contrast, its boundary critical behavior remains largely unexplored: it is unknown which microscopic boundary conditions flow to conformal boundary conditions, how many boundary fixed points exist, and what boundary conformal data characterize them.
This question is also relevant to quantum error correction, where decoding of surface codes maps to an RBIM-type statistical model and boundary or interface noise can affect decoding thresholds~\cite{dennis2002topological,wang2003confinement,ohno2004phase,bravyi2014efficient,fan2024diagnostics,Lee2025Exact,Sala2025Stability,Huang2025coherent,bravyi1998quantum,kitaev2003fault,horsman2012surface,fowler2012surface,ramette2023fault,Higgott2023improved,Marton2025Lattice}.

\begin{figure}
    \centering
    \includegraphics[width=1.0\linewidth]{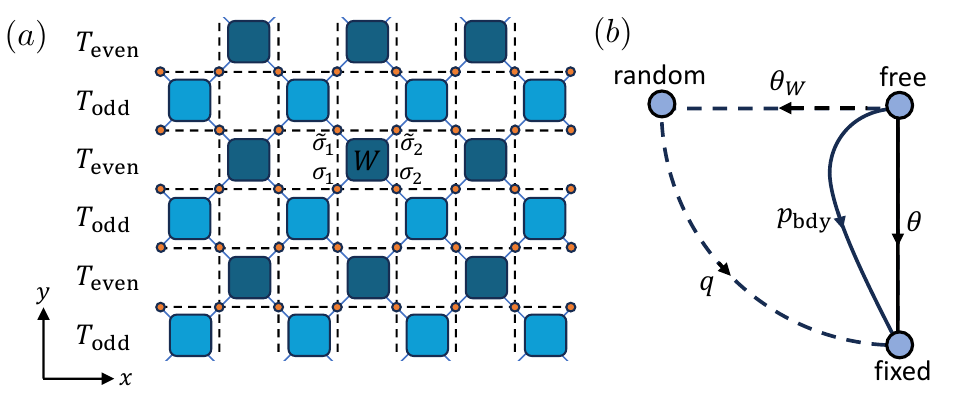}
    \caption{(a) Random-bond Ising model and its tensor-network representation. 
    Orange dots denote Ising spins, dashed lines denote random couplings, and each tensor $W$ encodes the Boltzmann weight of the four surrounding bonds. 
    The transfer matrices $T_{\rm odd}$ and $T_{\rm even}$ are built from alternating layers of $W$ tensors.
    (b) Schematic boundary RG flow among the free, fixed, and random conformal boundary conditions. 
    Solid black lines denote flows that preserve the Nishimori condition, while dashed lines indicate generic random flows that break it.}
    \label{fig:tn and RG flow}
\end{figure}

Here we develop a tensor-network approach for the RBIM at the Nishimori multicritical point and extract boundary conformal data; see Fig.~\ref{fig:tn and RG flow}. 
Using time-evolving block decimation method~\cite{vidal2003efficient,vidal2004efficient,daley2004time,cirac2004matrix,calabrese2004entanglement,zhou2006entanglement,calabrese2009entanglement,cardy2016entanglement}, we obtain the boundary entropy and identify three conformal boundary conditions: free, fixed, and random. 
Finite-size scaling shows that the free boundary is unstable, the fixed boundary is stable, and the random boundary lies on the flow between them. 
We further extract boundary-condition-changing (b.c.c.) operator dimensions from wavefunction overlaps~\cite{affleck1997boundary,zou2022universal,zhou2024the}, demonstrate boundary multifractality, and complement with a controlled $6-\epsilon$ boundary RG analysis of the replica field theory~\cite{le1988location,Le1989epsilon}. 
Finally, we relate the resulting BCFT data to boundaries of quantum error-correcting codes (QECC), where it diagnoses a boundary decoding transition of a planar surface code. 

\begin{figure*}
    \centering
    \includegraphics[width=1.0\linewidth]{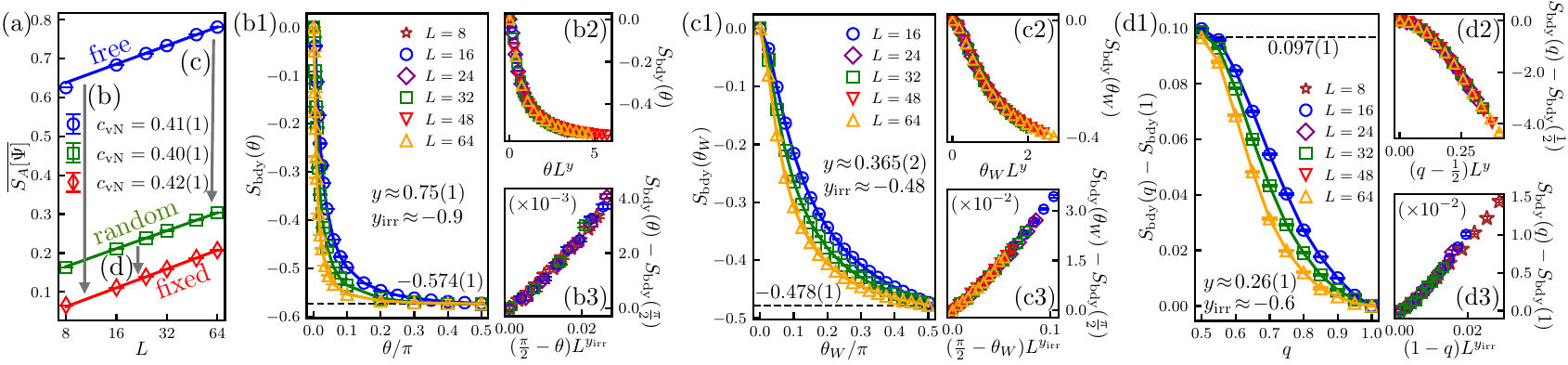}
    \caption{(a) Half-chain von Neumann entanglement entropy $\overline{S_{A}[\Psi]}$ versus $L$ evaluated at the free, random, and fixed boundary fixed points [denoted by circles in Fig.~\ref{fig:tn and RG flow}(b)], where the logarithmic-law fits yield the respective von Neumann central charges $c_\text{vN} \approx 0.41(1), 0.40(1),$ and $0.42(1)$\,. 
    Gray arrows indicate the corresponding boundary RG flows analyzed in (b1)-(b3), (c1)-(c3), and (d1)-(d3). (b1) Relative boundary entropy as a function of the spin-orientation angle $\theta$ for different $L$ in the disorder-free case ($q=1$), describing the boundary RG flow from the free to the fixed boundary condition. (b2), (b3) Data collapses near $\theta = 0$ and $\theta = \frac{\pi}{2}$, yielding the RG exponents, $y \approx 0.75(1)$ and $y_\text{irr} \approx -0.9$\,. (c1) Relative boundary entropy as a function of the boundary disorder strength $\theta_{W}$ for different $L$ with fixed $q = \frac{1}{2}$, describing the boundary RG flow from the free to the random boundary condition. (c2), (c3) Data collapses near $\theta_{W} = 0$ and $\theta_{W} = \frac{\pi}{2}$, yielding the RG exponents, $y \approx 0.365(2)$ and $y_\text{irr} \approx -0.48$\,. (d1) Relative boundary entropy as a function of the probability $q$ for different $L$ with fixed $\theta_{W} = \frac{\pi}{2}$, describing the boundary RG flow from the random to the fixed boundary condition. (d2), (d3) Data collapses near $q = \frac{1}{2}$ and $q = 1$, yielding the RG exponents, $y \approx 0.26(1)$ and $y_\text{irr} \approx -0.6$\,. In (b3), the sampling number is increased to $10^7$ to further suppress statistical errors. The $y$-axis values of (b3), (c3), and (d3) are rescaled by factors of $10^{3}$, $10^{2}$, and $10^{2}$, respectively, for better visual presentation.}
    \label{fig:g-function}
\end{figure*}

\paragraph{Model.}\ssll
The RBIM on a square lattice is defined by the partition function
\begin{equation}\label{eq:partition}
    Z = \sum_{\{\sigma\}} e^{\sum_{\langle ij \rangle} J_{ij} \sigma_i \sigma_j}\,, 
\end{equation}
where $\sigma_i = \pm 1$ denotes the spin at site $i$, $\left< ij \right>$ is the nearest-neighbor bond, and $J_{ij}$ is the bond coupling taken from $\pm J$ randomly with probability
\begin{equation}
    P(J_{ij}) =(1-p) \delta(J_{ij} - J) + p \, \delta(J_{ij} + J) \,, 
\end{equation}
where $0\le p \le 1$ is the probability for antiferromagnetic couplings.  
The Nishimori line is defined by the relation, $e^{-2J} = \frac{p}{1-p}$~\cite{nishimori1980exact,nishimori1981internal}. 
Along this line, there is a transition from ferromagnetic phase to paramagnetic phase at 
$p_c \approx 0.1092212(4)$~\cite{Wan2025NishimoriPoint}, corresponding to the Nishimori multicritical point.
We adopt this highly precise value of $p_c$ for all numerical calculations in this work.

The partition function can be represented by a tensor network as illustrated in Fig.~\ref{fig:tn and RG flow}(a).
The building block is $W_{\sigma_1,\sigma_2}^{\tilde \sigma_1, \tilde \sigma_2 } = e^{\sum_{\langle \sigma,\sigma' \rangle} J_{\sigma \sigma'} \sigma \sigma'} $ with $\sigma, \sigma' \in \{\sigma_1,\sigma_2, \tilde \sigma_1, \tilde \sigma_2 \}$, and $J_{\sigma \sigma'}$ the corresponding random bond coupling between spins $\sigma$ and $\sigma'$. 
It encodes the local Boltzmann weight associated with the four surrounding bonds. 
Using the transfer matrices, $T_{\rm odd}[\bm J] = \sum_{\bm \sigma,\tilde{\bm \sigma}} \prod_{i=1}^{L_x/2} W_{\sigma_{2i-1},\sigma_{2i}}^{\tilde\sigma_{2i-1},\tilde\sigma_{2i}} \left|  \tilde{\bm \sigma} \right> \left< \bm \sigma \right| $ and $T_{\rm even}[\bm J] = \sum_{\bm \sigma,\tilde{\bm \sigma}} \prod_{i=1}^{L_x/2}W_{\sigma_{2i},\sigma_{2i+1}}^{\tilde\sigma_{2i},\tilde\sigma_{2i+1}} \left|  \tilde{\bm \sigma} \right> \left< \bm \sigma \right|$, where $\bm J$ denotes the random bond variables contained in $W$ tensors, the partition function with the periodic boundary condition is given by $Z =\Tr\left[ \prod_{j=1}^{L_y/2} T_{\rm even}[\bm J_{2j} ] T_{\rm odd}[\bm J_{2j-1} ] \right]$~(see Supplemental Material for detail \cite{supplemental}).
Next, we will implement open boundary conditions with different boundary tensors.

\paragraph{Boundary entropy.}\ssll
We introduce the boundary tensor $B_\sigma^\theta = \cos(\frac{\pi}4 - \frac\theta2) \delta_{\sigma,+1} + \sin(\frac{\pi}4 - \frac\theta2) \delta_{\sigma,-1}$, where $\theta$ characterizes the boundary-spin orientation.
Besides the fixed boundary-spin orientation, at each boundary site, the orientation of the boundary tensor can be chosen randomly as $\pm\theta_W$ according to the distribution $P_{\theta_W,q}(\theta)=(1-q)\delta(\theta+\theta_W)+q\,\delta(\theta-\theta_W)$.
This construction gives microscopic boundary conditions that incorporate both boundary-spin rotation and boundary disorder.
We first consider the disorder-free case, $q=1$, where all boundary tensors have a fixed orientation $\theta$.
For example, defining $\left| B^\theta \right> = \sum_{\sigma} B_\sigma^\theta \left| \sigma \right>$, we obtain $\left< B^\theta \right| \vec \sigma \left| B^\theta \right> = \cos \theta \hat i + \sin \theta \hat k$.
In particular, $\theta = 0$ corresponds to the free boundary condition and $\theta = \pm \frac\pi2$ to fixed boundary conditions $\sigma = \pm 1$. 

With these boundary tensors, we define the open-boundary transfer matrices, $ T'_{\rm odd}[\bm J] = \sum_{\bm \sigma,\tilde{\bm \sigma}} B_{\tilde \sigma_1}^\theta B_{\sigma_1}^\theta \left(\prod_{i=1}^{L_x/2} W_{\sigma_{2i-1},\sigma_{2i}}^{\tilde\sigma_{2i-1},\tilde\sigma_{2i}} \right) B_{\tilde \sigma_{L_x}}^\theta B_{\sigma_{L_x}}^\theta \left|  \tilde{\bm \sigma}_{b} \right> \left< \bm \sigma_{b} \right|$ and $T'_{\rm even}[\bm J] = \sum_{\bm \sigma_b,\tilde{\bm \sigma}_b} \prod_{i=1}^{L_x/2-1}W_{\sigma_{2i},\sigma_{2i+1}}^{\tilde\sigma_{2i},\tilde\sigma_{2i+1}} \left|  \tilde{\bm \sigma}_{b} \right> \left< \bm \sigma_{b} \right|$, where $\bm{\sigma}_b = \{\sigma_2, \dots, \sigma_{L_x-1}\}$ denotes the bulk spins. 
They generate the state
\begin{equation}
    \label{eq:psi_theta}
    \left| \Psi_\theta(\bm{\mathrm{J}}) \right> = \frac1{\mathcal N'} F' \prod_{j=1}^{L_y/2} T'_{\rm even}[\bm J_{2j} ] T'_{\rm odd}[\bm J_{2j-1} ] \left|I \right> \,, 
\end{equation}
where $\mathcal N'$ is the normalization factor for the wavefunction and $F' = \sum_{\bm \sigma_b} e^{- \sum_{i=2}^{L_x-2} (-1)^i J_{\sigma_i\sigma_{i+1}} \sigma_i \sigma_{i+1}/2}  \left|  {\bm \sigma}_b \right> \left< \bm \sigma_b \right|$. 
This additional layer, $F'$, is introduced to ensure the correct evaluation of the von Neumann entropy and the corresponding wavefunction overlaps~\cite{supplemental}.

We then calculate the von Neumann entropy to reveal the boundary entropy~\footnote{We consider an initial state $\left|I\right>$ (the detail of which is not important) and evolve according to the transfer matrix for sufficiently large steps (equivalently, $L_y$).
To guarantee convergence, we have set $L_{y} = 10 L_{x}$ and kept the truncation error during the evolution below $10^{-10}$ across all simulations, ensuring that the results have reached their steady-state value with high accuracy. 
For simplicity, $L_{x}$ is also denoted as $L$ in the following discussion.}. 
Separating the system into two subsystems $A$ and $B$, the von Neumann entropy of the subsystem $A$ is defined as 
$S_{A} [\Psi_\theta(\bm{\mathrm{J}})] = - \Tr[\rho_A(\bm{\mathrm{J}}) \log \rho_A(\bm{\mathrm{J}}) ]$, where the reduced density matrix $\rho_A(\bm{\mathrm{J}}) = \Tr_{B} [ \left|\Psi_\theta(\bm{\mathrm{J}}) \right> \left<\Psi_\theta(\bm{\mathrm{J}}) \right| ]$. 
We show in the End Matter that the half-chain (i.e., $L_A = L/2$) von Neumann entropy gives~\cite{appendix}
\begin{equation} \label{eq:vN_entropy_main}
\begin{split}
    \overline{S_{A}[\Psi_\theta]} 
     = \frac{c_{\rm vN}}6  \log \frac{L}\pi + \tilde S_{\rm bdy}(\theta) + s_{\rm UV}   \,,
\end{split}
\end{equation}
where $\overline{(\cdot)}=\sum_{\bm{\mathrm{J}}} P(\bm{\mathrm{J}})(\cdot)$, $P(\bm{\mathrm{J}})$ is the probability of the random bond configuration $\bm{\mathrm{J}}$.
If the boundary tensor is chosen randomly, the disorder average $\overline{(\cdot)}$ is understood to include the average over the boundary-tensor ensemble as well.
$c_{\rm vN}$ is the central charge defined via the von Neumann entropy, $\tilde S_{\rm bdy}(\theta)$ is the absolute boundary entropy, and $s_{\rm UV}$ is a non-universal constant. 
Note that in nonunitary CFTs, $c_{\rm vN}$ need not coincide with the central charge $c$ of the Virasoro algebra. 
In the main text, we focus on the boundary entropy difference relative to the free boundary condition,
$S_{\rm bdy}(\theta) \equiv \tilde S_{\rm bdy}(\theta)-\tilde S_{\rm bdy}(0) =\overline{S_A[\Psi_\theta]} -
\overline{S_A[\Psi_{\theta=0}]}$.
The absolute boundary entropy $\tilde S_{\rm bdy}(0)$ is determined
independently in the Supplemental Material~\cite{supplemental}. 
The results for the von Neumann entropy are shown in Fig.~\ref{fig:g-function}(a).
The disorder average is typically performed over $10^6$ samples in this work.  
Fig.~\ref{fig:g-function}(a) agrees with Eq.~\eqref{eq:vN_entropy_main} with fitted central charge $c_{\rm vN} \approx 0.42$, consistent with Ref.~\cite{Putz2025NishimoriCharge}. 

The $S_{\rm bdy}(\theta)$ is a monotonically decreasing function for $\theta \in (0, \pi/2)$ as plotted in Fig.~\ref{fig:g-function}(b1). 
This alone does not establish conformal boundary conditions, since the $g$-theorem need not hold in nonunitary CFTs. 
We therefore analyze the finite-size scaling near the candidate fixed points, $S_{\rm bdy}(\theta_\ast+\delta\theta)-S_{\rm bdy}(\theta_\ast) \simeq f(\delta\theta L^{y})$, where $\theta_\ast=0$ and $\theta_\ast=\pi/2$ correspond to the free and fixed boundaries, respectively. 
The collapse in Fig.~\ref{fig:g-function}(b2) shows that the free boundary is an unstable conformal boundary condition, with a relevant perturbation of exponent $y\simeq0.75(1)$; by contrast, Fig.~\ref{fig:g-function}(b3) shows that the fixed boundary is stable, with leading irrelevant exponent $y_{\rm irr}\simeq -0.9$\,. 
This identifies a boundary RG flow from the free to the fixed conformal boundary condition, as summarized in Fig.~\ref{fig:tn and RG flow}(b).

We next consider a second subspace of boundary states, characterized by a random boundary-spin orientation.
We set $q=1/2$ and choose the boundary-spin orientation to be $\theta=\pm\theta_W$, with $\theta_W\in[0,\frac{\pi}{2}]$.
Equivalently, each boundary tensor is independently oriented as $\pm\theta_W$ with equal probability.
Note that in the limit, $\theta_W=\pi/2$, this construction realizes a strong random boundary field.
While such a random boundary field is marginally irrelevant for the clean 2D Ising model~\cite{Cardy1991TheIsing,Pleimling2004Logarithmic}, we find that it becomes relevant at the Nishimori multicritical point.
As shown in Fig.~\ref{fig:g-function}(c), increasing the system size drives the boundary entropy toward a new stable value.
Finite-size collapses further show that the disorder perturbation is relevant at the free boundary, establishing a distinct disorder-dominated conformal boundary condition, which we denote as the random boundary fixed point. 

We further study the boundary RG flow between the random boundary fixed point and the fixed boundary fixed point.
This flow is realized by fixing $\theta_W = \pi/2$ and tuning the probability from $q=\frac{1}{2}$ to $q=1$.
Although the random boundary fixed point is stable against tuning $\theta_W$, we find that it is unstable along this probability-tuning direction and flows to the fixed boundary fixed point, as shown in Fig.~\ref{fig:g-function}(d).
This flow also shows that the boundary RG trajectory runs from a fixed point with larger boundary entropy to one with smaller boundary entropy.
From the finite-size data collapse, we extract the relevant perturbation exponent near the random boundary fixed point, as well as the leading irrelevant perturbation exponent near the fixed boundary fixed point.

Together with the free, fixed, and random boundary fixed points, and the RG flows connecting them, these results complete the boundary RG structure shown in Fig.~\ref{fig:tn and RG flow}(b).
The existence of the random boundary fixed point demonstrates that Nishimori criticality supports boundary universality classes that are absent in the clean Ising model.

\paragraph{b.c.c. operators.}\ssll 
After identifying three conformal boundary conditions, we are in a position to extract the scaling dimension of the b.c.c.\ operator~\footnote{We provide strong numerical evidence for the conformal symmetry, see SM.}. 
For b.c.c.\ operator between boundary conditions $\alpha=a,\, b$, we consider the steady states $\left|\Psi_\alpha(\bm{\mathrm{J}}_i) \right>$, where $\bm{\mathrm{J}}_i$ are independent random bond couplings.
Then the b.c.c.\ scaling dimension is encoded in the wavefunction overlap ratio~(see End Matter for detail \cite{appendix}), 
\begin{eqnarray} \label{eq:bcc}
    && \overline{\log Q_{ab}} \simeq -2\Delta_{ab} \log L , \, \\
    && Q_{ab} = \sqrt{\frac{\left< \Psi_a(\bm{\mathrm{J}}_1)| \Psi_b(\bm{\mathrm{J}}_2) \right> \left< \Psi_a(\bm{\mathrm{J}}_2)| \Psi_b(\bm{\mathrm{J}}_1) \right>}{\left< \Psi_a(\bm{\mathrm{J}}_1)| \Psi_a(\bm{\mathrm{J}}_2) \right> \left< \Psi_b(\bm{\mathrm{J}}_2)| \Psi_b(\bm{\mathrm{J}}_1) \right>}} \,, \label{eq:ratio}
\end{eqnarray}
where $\Delta_{ab}$ is the typical scaling dimension of the b.c.c.\ operator.
The ratio is to properly cancel the normalization factor in each wavefunction. 

\begin{figure}
    \centering
    \includegraphics[width=1.0\linewidth]{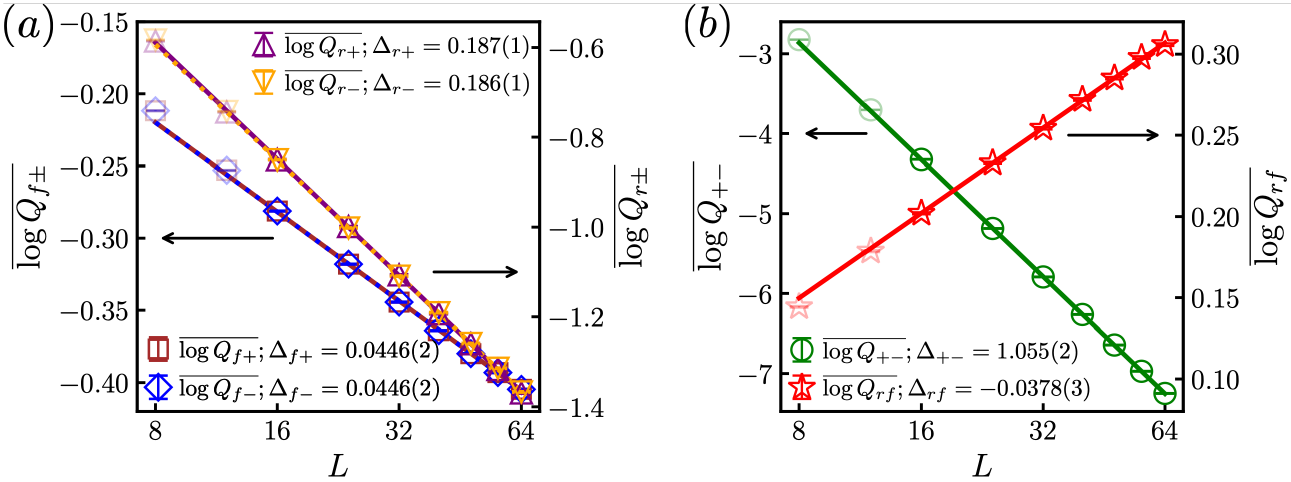}
    \caption{
    Wavefunction overlaps versus lattice size $L$. (a) The least-squares fittings for $Q_{f\pm}$ and $Q_{r\pm}$. (b) The least-squares fittings for $Q_{rf}$ and $Q_{+-}$. According to $\overline{\log Q_{ab}} \simeq -2\Delta_{ab} \log L$, the estimated scaling dimensions of the b.c.c. operators are $\Delta_{f\pm} \approx 0.0446(2)$, $\Delta_{r\pm} \approx 0.186(2)$, $\Delta_{rf} \approx -0.0378(3)$, and $\Delta_{+-} \approx 1.055(2)$. The subscripts $f$, $\pm$, and $r$ denote the free, fixed $\pm$, and random boundary states, respectively. The two data points with the smallest values of $L$ were excluded from the fitting.}
    \label{fig:bcc-main}
\end{figure}

For convenience, we term the boundary condition of the first family for $\theta=  0 $ as $f$ (free) boundary condition and $\theta = \pm \frac\pi2$ as $\pm$ (fixed) boundary conditions. 
We also denote the disorder-induced random conformal boundary condition by $r$.
In Fig.~\ref{fig:bcc-main}, we plot the wavefunction overlap of $\overline{\log Q_{f\pm}} $ and $\overline{\log Q_{r\pm}} $ in panel (a), and $\overline{\log Q_{+-}}$ and $\overline{\log Q_{rf}}$ in panel (b). 
The data follow the scaling form in Eq.~\eqref{eq:bcc} well, yielding
$\Delta_{f\pm}\approx 0.0446(2)$,
$\Delta_{r\pm}\approx 0.186(2)$,
$\Delta_{+-}\approx 1.055(2)$,
and $\Delta_{rf}\approx -0.0378(3)$.

\paragraph{Boundary multifractality.}\ssll
We continue to investigate the boundary multifractality for the free boundary condition. 
Distinct from the previous calculation, we consider a periodic boundary condition for the transfer matrix, and utilize a ``spacetime'' rotation. 
Then, the free boundary condition is characterized by the state 
$\left|X \right> = \otimes_{i=1}^{L_x} \frac1{\sqrt2} \sum_{\sigma_i = \pm 1}\left| \sigma_i \right> $. 
The boundary spin-spin correlation function can be expressed as  
\begin{equation}
\label{eq:psi_spin}
\begin{split}
    & \left< \sigma_i \sigma_j \right> = \frac{\left< X | \sigma^z_i \sigma^z_j | \Psi(\bm{\mathrm{J}}) \right>}{\left< X  | \Psi(\bm{\mathrm{J}}) \right>} \,, \\
    & \left| \Psi(\bm{\mathrm{J}}) \right> =  \frac1{\mathcal{N}} F \prod_{j=1}^{L_y/2} T_{\rm even}[\bm J_{2j} ] T_{\rm odd}[\bm J_{2j-1} ] \left| I \right>\,, 
\end{split}
\end{equation}
where $\mathcal N$ is the normalization factor for the wavefunction, and the additional transfer matrix, $F =  \sum_{\bm \sigma} e^{\sum_{i=1}^{L_x/2} J_{\sigma_{2i-1}\sigma_{2i}} \sigma_{2i-1} \sigma_{2i}}  \left|  {\bm \sigma} \right> \left< \bm \sigma \right|$, is to compensate for the bonds missed in the brick-wall pattern for the last layer.

\begin{figure}
    \centering
    \includegraphics[width=0.6\linewidth]{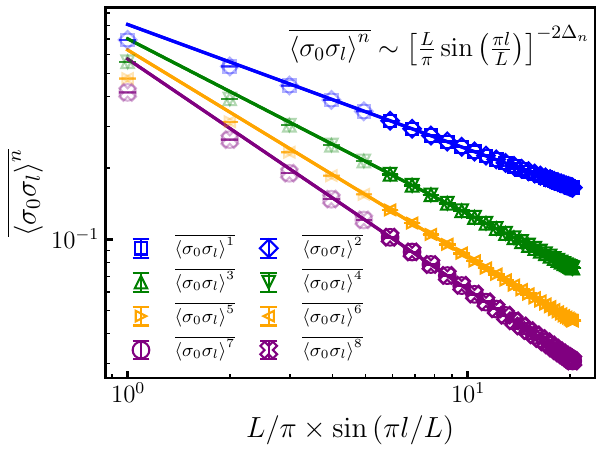}
    \caption{Moments of the boundary spin-spin correlation function, $\overline{\left< \sigma_0 \sigma_l \right>^n}$, versus the conformal chord length $L/\pi \sin(\pi l/L)$ for $n=1$ to $8$ with size $L = 64$. The power-law fittings according to the conformal ansatz Eq.~\eqref{eq:cor_scaling} estimate the corresponding exponents as $\Delta_{1} = \Delta_{2} \approx 0.263(1)$, $\Delta_{3} = \Delta_{4} \approx 0.369(1)$, $\Delta_{5} = \Delta_{6} \approx 0.436(1)$, and $\Delta_{7} = \Delta_{8} \approx 0.486(1)$\,. The fits are performed by excluding several data points with relatively small $l$ (indicated by lighter colored markers).}
    \label{fig:correlation}
\end{figure}

We compute the $n$-th moment of the boundary spin-spin correlation function,
$\overline{\langle \sigma_0\sigma_l\rangle^n}$, where the site $0$ is an arbitrary reference point. 
The results are shown in Fig.~\ref{fig:correlation}. 
For the free boundary condition, the Nishimori gauge symmetry implies the exact relation
$\overline{\langle\sigma_i\sigma_j\rangle^{2k-1}}
=
\overline{\langle\sigma_i\sigma_j\rangle^{2k}}$
for boundary spin correlations~\cite{nishimori1980exact,nishimori1981internal}, which is confirmed by our data. 
The moments obey the conformal scaling form
\begin{equation}
    \label{eq:cor_scaling}
    \overline{\left<\sigma_0 \sigma_l \right>^{n}} \simeq \left( \frac{L}\pi \sin\left(\frac{\pi l}{L} \right) \right)^{-2\Delta_n} \,,
\end{equation}
with fitted exponents given in Fig.~\ref{fig:correlation}. 
Interestingly, the scaling dimension of the spin operator $\Delta_1 = \Delta_2$ also governs the relevant exponents that drive the boundary RG flow away from the free boundary fixed point as discussed in Sec.~V in the Supplemental Material~\cite{supplemental}.

\paragraph{RG calculation.}\ssll
To complement our nonperturbative BCFT data, we present a controlled boundary RG analysis of the replicated Landau theory  for the RBIM near its upper critical dimension $d_c=6$, and focus on the Dirichlet boundary condition~\cite{diehl1986field,diehl1996the} 
that corresponds to the natural candidate for the free boundary condition studied above.  
We find that, at one loop in $\epsilon\equiv 6-d$, the $n$-th moment of the boundary spin-spin correlation has the multifractal dimension~\cite{supplemental},
\begin{equation}
\Delta_{2k-1}(d) = \Delta_{2k}(d) = \frac{kd}{2}+\frac{k(2-5k)}{3} \epsilon + \mathcal O(\epsilon^2) \,.
\label{eq:DeltaMp_eps}
\end{equation}

The raw $\epsilon$-expansion is quantitatively unreliable when extrapolated to $d=2$, but it still provides
controlled information near $d=6$ and, importantly, fixes the analytic structure needed for resummation. 
Our numerical extraction of the boundary-spin scaling dimension for the free boundary condition supplies a nonperturbative anchor in $d=2$.
As a concrete example, we focus on $n=1$, combining $\Delta_{1}(6)=3$ and
$\partial_\epsilon \Delta_1|_{\epsilon=0}= - \frac{3}{2}$ with the $d=2$ input $\Delta_{1}(2)\approx 0.263$ to construct a minimal $[1/1]$ Pad\'e approximation~\cite{Baker1961The,diehl1989semi}, $
\Delta^{[1/1]}_1(d)=\frac{3-0.606\,\epsilon}{1+0.298\,\epsilon}$. 
While we do not claim precision from this simple scheme, it illustrates how boundary conformal data in $d=2$ can be used to constrain resummations of the controlled $6-\epsilon$ expansion and
provides a systematic bridge between perturbative RG and the BCFT numerics.

\begin{figure}
    \centering
    \includegraphics[width=1.0\linewidth]{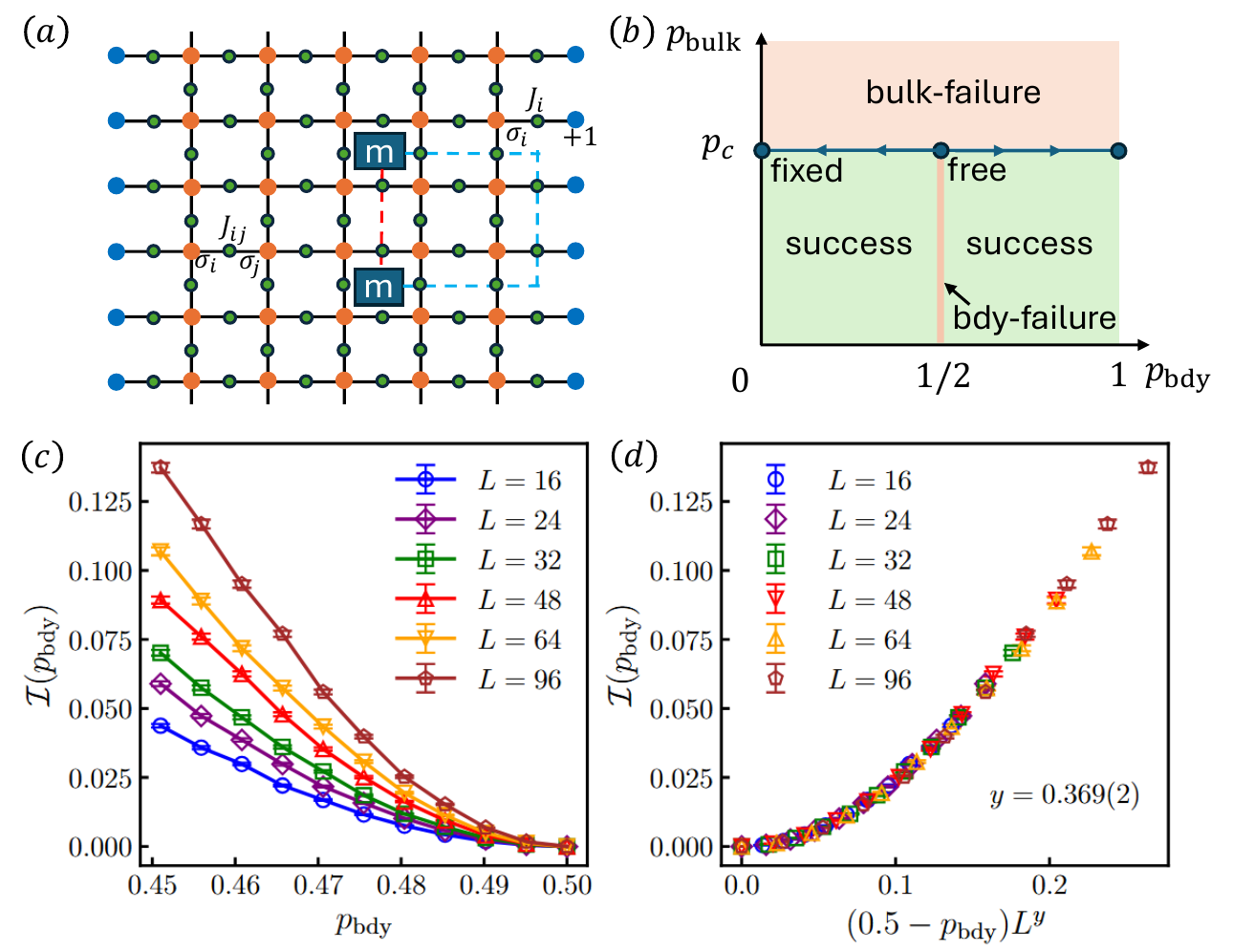}
    \caption{(a) Surface code with rough boundaries in $x$ direction and periodic boundary conditions in $y$.  
    The qubits live on the edges denoted by green dots.
    The corresponding bulk spins of RBIM live on the vertices denoted by orange dots.
    Red and blue dashed lines are examples of error chains. 
    The $m$ particle cannot be absorbed by the rough boundary, leading to the outermost spins pinning at $+1$, illustrated by the blue dots.  
    (b) Bulk and boundary error threshold separating the decoding success phase and failure phase. 
    Their intersection corresponds to the free boundary universality class. 
    $p_{\rm bdy} < \frac12$ ($p_{\rm bdy} > \frac12$) drives the RG flow from the free to fixed boundary universality class.  
    (c),(d) Coherent information $\mathcal{I}$ on a cylinder versus boundary error rate $p_{\rm bdy}$, with bulk error fixed at $p=p_c$. 
    In the RBIM, $p_{\rm bdy}$ controls the coupling between boundary spins and fixed exterior spins. 
    (c) $\mathcal{I}(p_{\rm bdy})$ for different $L$. 
    (d) Data collapse near $p_{\rm bdy}=1/2$, giving $y=0.369(2)$.}
    \label{fig:QECC}
\end{figure}

\paragraph{Boundary of QECC.}\ssll
The Nishimori point has a direct interpretation in QECC. 
For the toric code with plaquette stabilizers $B_p=\prod_{i\in p} Z_i$ and vertex stabilizers $A_v=\prod_{i\in v} X_i$, independent bit-flip errors $\mathcal{E}_i(\rho)=(1-p)\rho+pX_i\rho X_i$ map bulk decoding to the RBIM on the Nishimori line, leading to the bulk decoding threshold at the Nishimori multicritical point. 
For a surface-code cylinder with rough boundaries in one direction and periodic boundary conditions in the other as illustrated in Fig.~\ref{fig:QECC}(a), the open boundaries introduce an additional boundary threshold: in the boundary-only limit, this channel leads to a boundary threshold, $p_{\rm bdy,c}=1/2$~\cite{appendix}.
At the intersection of the bulk decoding critical point and the boundary threshold, see Fig.~\ref{fig:QECC}(b), $(p_{\rm bulk},p_{\rm bdy})=(p_c,1/2)$, one is naturally led to ask whether the boundary of the Nishimori CFT flows to a conformal boundary condition~\footnote{In the two-dimensional surface-code geometry considered here, boundary error correction fails only at $p_{\rm bdy,c}=1/2$, whereas in higher-dimensional codes the boundary error threshold may lie below $1/2$~\cite{tan2024resilience,ramette2023fault}.}. 

For a rough boundary, the plaquette operator is truncated at the edge of
the surface code. Since an $m$ particle cannot be absorbed by this
boundary, an $X$-error string cannot terminate there: the domain wall
connecting the corresponding $m$ syndromes must remain in the bulk, see Fig.~\ref{fig:QECC}(a). In the corresponding RBIM model, this constraint is implemented by fixing the outermost Ising spins, which we choose to be $+1$ by a gauge choice.
The boundary layer next to the outermost is therefore described by the one-body random
boundary term
\begin{equation} \label{eq:bdy_qec}
    H_{\rm bdy}=-\sum_{i\in {\rm bdy}} J_i \sigma_i \,,
\end{equation}
with boundary disorder satisfying the Nishimori relation $e^{-2J_{\rm bdy}}=\frac{p_{\rm bdy}}{1-p_{\rm bdy}}$.
$p_{\rm bdy}=0$ gives $J_{\rm bdy}\to\infty$, which pins the boundary spins and realizes the fixed boundary condition. 
$p_{\rm bdy}=1/2$ gives $J_{\rm bdy}=0$, so the boundary field vanishes and the boundary layer is left free. Tuning $p_{\rm bdy}$ away from $1/2$ therefore introduces a Nishimori-symmetric boundary perturbation that drives the RG flow from the free to the fixed boundary fixed point, as shown in Fig.~\ref{fig:tn and RG flow}(b). 

We diagnose the boundary threshold at the Nishimori multicritical point using the coherent information~\cite{Wan2025NishimoriPoint} $\mathcal{I}(p_{\rm bdy})= 1+\overline{\log_2\left[ \frac{Z(p_{\rm bdy})}   {Z(p_{\rm bdy})+Z'(p_{\rm bdy})} \right]}$, where $Z$ and $Z'$ are the RBIM partition functions in the sectors with identical and opposite fixed boundary conditions~\cite{supplemental}. 
As shown in Fig.~\ref{fig:QECC}(c), $\mathcal{I}$ collapses at $p_{\rm bdy}=\frac12$, indicating an unstable boundary fixed point. 
The collapse in Fig.~\ref{fig:QECC}(d) gives the relevant exponent
$y\approx0.369(2)$.
Note that this exponent is consistent with the relevant exponent from the quenched random boundary field perturbation in Fig.~\ref{fig:g-function}(c).
With the boundary noise as an effective random boundary field as shown in Eq.~\eqref{eq:bdy_qec} that locally pins the boundary spins, tuning $p_{\rm bdy}\lesssim \frac12$ and tuning the boundary randomness $\theta_W\gtrsim0$ couple to the same leading relevant boundary operator, explaining the consistent numerical values of their scaling exponents. 
They nevertheless define distinct RG trajectories, since tuning $p_{\rm bdy}$ preserves the Nishimori gauge symmetry, whereas a generic random boundary orientation breaks it, as illustrated in Fig.~\ref{fig:tn and RG flow}(b).  
The BCFT data provide a universal description of boundary decoding phenomena in QECCs.

\paragraph{Conclusion.}\ssll 
Boundary universality at disordered or nonunitary critical points, exemplified by the Nishimori point, remains conceptually rich yet underexplored.  
Our work provides the first systematic BCFT characterization of Nishimori boundary criticality, identifying its conformal boundary conditions and associated boundary conformal data, and connects them to boundary decoding transitions in surface codes.
Looking forward, it would be valuable to generalize this framework to interfaces, corners, and disordered boundary ensembles, and to explore possible generalizations of the $g$-theorem at the Nishimori point~\cite{patil2025shannon}.

{\it Acknowledgments:} We thank Shuo Liu and Zhou-Quan Wan for helpful discussions.
Numerical simulations were carried out with the ITENSOR \verb|C++| package~\cite{itensor}. 
This work is supported in part by a start-up fund (S.-K.~J.) at Tulane University. 
The work of S.Y. is supported by China Postdoctoral Science Foundation (Certificate Number: 2024M752760). S.Y. would like to thank Tulane University for its hospitality during the visit when this work was initiated.

\bibliography{reference.bib}

\appendix
%\onecolumngrid
\section{\large{End Matter}}
%\twocolumngrid

\paragraph{von Neumann entropy.}\ssll
In this section, we derive Eq.~\eqref{eq:vN_entropy_main} from the path-integral construction and the conformal symmetry.
For a fixed disorder realization $\bm{\mathrm{J}}$, the reduced density matrix of subsystem $A$ is $\rho_A(\bm{\mathrm{J}})=\mathrm{Tr}_B |\Psi_\theta(\bm{\mathrm{J}})\rangle \langle \Psi_\theta(\bm{\mathrm{J}})|$.  
The standard replica construction gives
\begin{equation}
    \mathrm{Tr}_A \left[ \rho_A(\bm{\mathrm{J}})^n \right] = \frac{Z_n(A;\theta,\bm{\mathrm{J}})}{Z_1(\theta,\bm{\mathrm{J}})^n} \,,
\end{equation}
where $Z_1(\theta,\bm{\mathrm{J}})$ is the single-copy partition function with boundary spin orientation $\theta$, and $Z_n(A;\theta,\bm{\mathrm{J}})$ is the partition function of an $n$-sheeted geometry obtained by cyclically gluing the $n$ copies along the region $A$.  
In the tensor-network language, this is obtained by stacking $n$ copies of the network for $|\Psi_\theta(\bm{\mathrm{J}})\rangle\langle\Psi_\theta(\bm{\mathrm{J}})|$ and inserting a cyclic permutation of the physical indices along $A$.  
The von Neumann entropy is then
\begin{equation} \label{eq:derivative_S1}
    S_A[\Psi_\theta(\bm{\mathrm{J}})] = -\partial_n\log \frac{Z_n(A;\theta,\bm{\mathrm{J}})}{Z_1(\theta,\bm{\mathrm{J}})^n} \bigg|_{n=1} \,.
\end{equation}

The only universal input needed in the following step is conformal covariance of the scaling limit.  
In a clean unitary CFT, this is the familiar Calabrese--Cardy argument~\cite{calabrese2009entanglement,cardy2016entanglement}.  
However, unitarity is not essential for the form of the result: conformal covariance of primary fields continue to fix the position dependence of correlation functions in a nonunitary CFT.  
At the Nishimori multicritical point, the disorder-averaged theory is expected to be a nonunitary CFT.  
The conformal symmetry at Nishimori multicritical point is further supported numerically in the Supplemental Material~\cite{supplemental}.  
Therefore, the cyclic gluing along $A$ can still be represented, in the scaling limit, by the insertion of a twist operator ${\cal T}_n$ at the entanglement cut:  
\begin{eqnarray}
    \overline{\left[\frac{Z_n(A;\theta,\bm{\mathrm{J}})}{Z_1(\theta,\bm{\mathrm{J}})^n}\right]^q} = \overline{\langle \mathcal T_n \rangle_\theta^q} \,,
\end{eqnarray}
where $\overline{(\cdot)}$ is the probability of the random bond configuration $\bm{\mathrm{J}}$ and if the boundary tensor is chosen randomly, the disorder average $\overline{(\cdot)}$ is understood to include the average over the boundary-tensor ensemble as well.
$\langle \mathcal T_n \rangle_\theta$ denotes the expectation value of the twist operator for an interval ending on a conformal boundary under the boundary condition $\theta$ in a fixed disorder realization.  
An additional copy, $q$, is introduced to use the following trick 
\begin{eqnarray}
    \overline{\log \langle \mathcal T_n \rangle_\theta} = \partial_q \overline{\langle \mathcal T_n \rangle^q_\theta} \,\Big|_{q=0} = \partial_q\log \overline{\langle \mathcal T_n \rangle^q_\theta} \,\Big|_{q=0}  \,,
\end{eqnarray}
where we used $\overline{\langle \mathcal T_n \rangle^q_\theta }\,\big|_{q=0} = 1$.

Utilizing the conformal symmetry, the one-point function of the twist operator in the disorder-average replica theory under boundary condition $\theta$ is constrained as 
\begin{eqnarray}
    \overline{\langle \mathcal T_n \rangle_\theta^q} = A_{n,q}(\theta) \left(\frac{L}{\pi a} \sin \left(\frac{\pi L_A}{L}\right)\right)^{- x_{n,q}}\,,
\end{eqnarray} 
where $L$ is the total length, $L_A$ is the length of the interval $A$, $a$ is a short-distance cutoff, $A_{n,q}$ is an $L$-independent boundary amplitude, and $x_{n,q}$ is the scaling dimension of the twist operator $\mathcal T_n$.  
Taking $L_A = L/2$, the von Neumann entropy is 
\begin{eqnarray}
    \overline{S_A[\Psi_\theta]} &=& - \partial_n \overline{ \log\langle \mathcal T_n \rangle_\theta}\Big|_{n=1} \\
    &=& - \partial_n \left[ \partial_q\log \overline{\langle \mathcal T_n \rangle^q_\theta} \,\Big|_{q=0}\right] \bigg|_{n=1} \\
    &=& \frac{c_{\rm vN}}6  \log \frac{L}\pi + \tilde S_{\rm bdy}(\theta) + s_{\rm UV} \,,
\end{eqnarray}
where the von Neumann central charge is $\frac{c_{\rm vN}}6 = \partial_{n} \partial_q x_{n,q} \Big|_{n=1,q=0}$, and the boundary entropy is $\tilde S_{\rm bdy}(\theta) = - \partial_{n} \partial_q \log A_{n,q}(\theta) \Big|_{n=1,q=0}$, and $s_{\rm UV}$ is related to the nonuniversal constant $a$.
This is Eq.~\eqref{eq:vN_entropy_main}.  
Thus the extension from the clean CFT formula to the Nishimori multicritical point relies only on the conformal symmetry, not on unitarity.  

\begin{figure}
    \centering
    \includegraphics[width=1\linewidth]{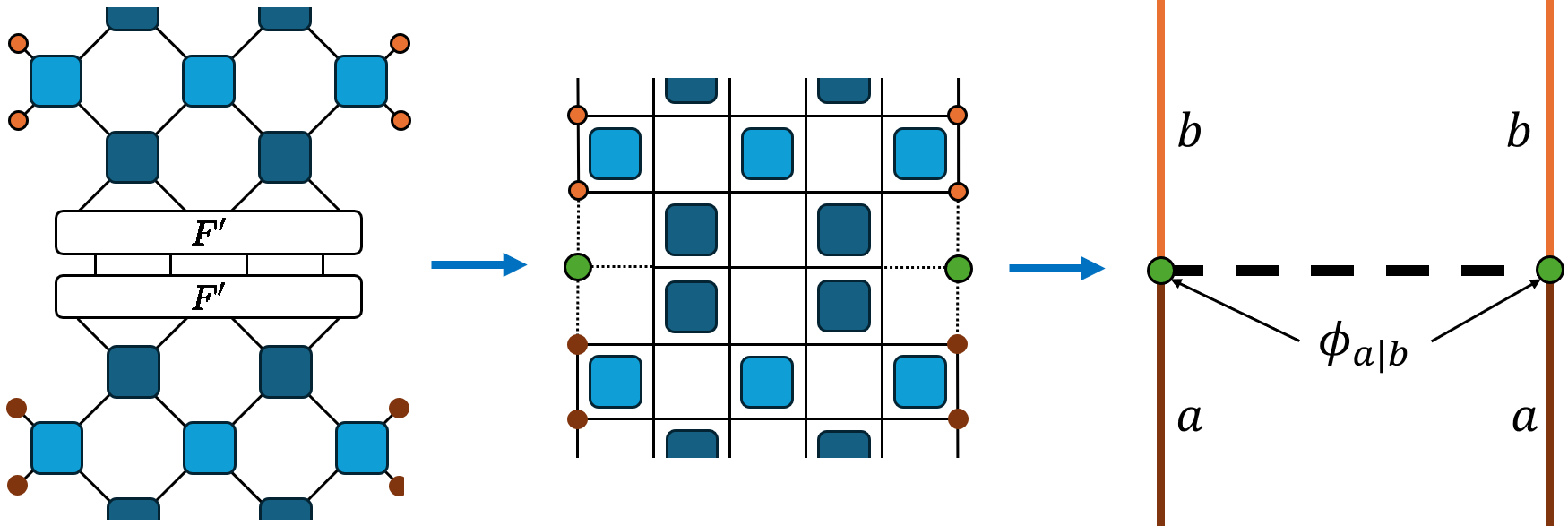}
    \caption{Left: Tensor-network representation of wavefunction overlap $\left< \Psi_{\theta'} | \Psi_\theta \right>$.
    The orange and brown dots denote the boundary spin orientations $\theta'$ and $\theta$, respectively. 
    Middle: Illustration of partition function, obtained from the wavefunction overlap~\cite{supplemental}.
    The green dots only indicate the insertion position of b.c.c.\ operators.
    Right: Insertion of b.c.c.\ operators in the BCFT description. 
    The b.c.c.\ operators $\phi_{a|b}$ are indicated by the green dots. }
    \label{fig:append_bcc}
\end{figure}

\paragraph{Wavefunction overlap.}\ssll
In this section, we relate the wavefunction overlap to the scaling dimension of the b.c.c. operator.
The wavefunction overlap $\langle \Psi_{\theta'}(\bm{\mathrm{J}}_2) | \Psi_{\theta}(\bm{\mathrm{J}}_1) \rangle$, as illustrated in the left panel of Fig.~\ref{fig:append_bcc}, is related to the partition function, $\langle \Psi_{\theta'}(\bm{\mathrm{J}}_2) | \Psi_{\theta}(\bm{\mathrm{J}}_1) \rangle \propto Z_{\theta',\theta}(\bm{\mathrm{J}}_2,\bm{\mathrm{J}}_1)$.
Here $Z_{\theta',\theta}(\bm{\mathrm{J}}_2,\bm{\mathrm{J}}_1)$ denotes the partition function on the sewn geometry in which the upper and lower halves carry disorder realizations $\bm{\mathrm{J}}_2$ and $\bm{\mathrm{J}}_1$, respectively, with boundary conditions $\theta'$ and $\theta$ imposed on the corresponding outer boundaries. 

In the scaling limit, the wavefunction overlap $\langle \Psi_{\theta'}(\bm{\mathrm{J}}_2) | \Psi_{\theta}(\bm{\mathrm{J}}_1) \rangle$ admits a natural BCFT interpretation in terms of b.c.c.\ operators, as illustrated in Fig.~\ref{fig:append_bcc}. 
Let us denote by $a$ and $b$ the conformal boundary conditions.
The geometry of the overlap is that of a long strip with boundary condition $a$ at the ``bottom'' and $b$ at the ``top'' (see Fig.~\ref{fig:append_bcc} middle and right panel). 
There are two points where the bottom and top boundaries meet; at these points, the boundary condition jumps from $a$ to $b$. 
In the BCFT description, these junctions are precisely the insertion points of b.c.c.\ operators $\phi_{a|b}$, which we indicate schematically by green dots in Fig.~\ref{fig:append_bcc}.

However, we should be careful about the normalization factor of the wavefunction.
Since the wavefunction is normalized, $\langle \Psi_{a}(\bm{\mathrm{J}}_1) | \Psi_{a}(\bm{\mathrm{J}}_1) \rangle =1$, the wavefunction overlap gives
\begin{eqnarray}
    \langle \Psi_{a}(\bm{\mathrm{J}}_1) | \Psi_{b}(\bm{\mathrm{J}}_2) \rangle = \frac{Z_{ab}(\bm{\mathrm{J}}_1, \bm{\mathrm{J}}_2)}{\sqrt{Z_{aa}(\bm{\mathrm{J}}_1, \bm{\mathrm{J}}_1) Z_{bb}(\bm{\mathrm{J}}_2, \bm{\mathrm{J}}_2)}} \,.
\end{eqnarray}
Note that the denominator in this expression involves nonstandard partition functions with correlated disorder configurations.  
For example, $Z_{aa}(\bm{\mathrm{J}},\bm{\mathrm{J}})$ denotes the partition function in which the disorder realization in the upper half is taken
to be identical to that in the lower half.

This is different from the standard b.c.c. correlator under a fixed configuration.  
For a fixed sewn disorder geometry, with independent disorder realizations $\bm{\mathrm{J}}_2$ and $\bm{\mathrm{J}}_1$ in
the upper and lower halves, respectively, the b.c.c. two-point function is
\begin{eqnarray}
    \left< \phi_{a|b}(0) \phi_{a|b}(L) \right>_{\bm{\mathrm{J}}_1,\bm{\mathrm{J}}_2}= \frac{ Z_{ab}(\bm{\mathrm{J}}_1,\bm{\mathrm{J}}_2)}{\sqrt{Z_{aa}(\bm{\mathrm{J}}_1,\bm{\mathrm{J}}_2) Z_{bb}(\bm{\mathrm{J}}_1,\bm{\mathrm{J}}_2)}}\, ,
\end{eqnarray}
where $\left< \cdot \right>_{\bm{\mathrm{J}}_1,\bm{\mathrm{J}}_2}$ denotes the expectation value under the disorder realization: $\bm{\mathrm{J}}_1$, $\bm{\mathrm{J}}_2$ for the upper and lower halves, respectively.   
Importantly the same disorder geometry is used in the numerator and
denominator, while only the boundary conditions are changed.

This motivates the ratio defined in Eq.~\eqref{eq:ratio}, which is equivalent to the following expression, 
\begin{eqnarray}
    Q_{ab} =\sqrt \frac{Z_{ab}(\bm{\mathrm{J}}_1, \bm{\mathrm{J}}_2) Z_{ab}(\bm{\mathrm{J}}_2, \bm{\mathrm{J}}_1)}{{Z_{aa}(\bm{\mathrm{J}}_1, \bm{\mathrm{J}}_2) Z_{bb}(\bm{\mathrm{J}}_1, \bm{\mathrm{J}}_2)}} \,.
\end{eqnarray}
Now by taking logarithm, we have
\begin{eqnarray}
    \log Q_{ab} &=& \frac12 \Big[ \log \frac{Z_{ab}(\bm{\mathrm{J}}_1,\bm{\mathrm{J}}_2)}{\sqrt{Z_{aa}(\bm{\mathrm{J}}_1,\bm{\mathrm{J}}_2) Z_{bb}(\bm{\mathrm{J}}_1,\bm{\mathrm{J}}_2)}} \nonumber \\
    && + \log \frac{
    Z_{ab}(\bm{\mathrm{J}}_2,\bm{\mathrm{J}}_1)
    }{
    \sqrt{
    Z_{aa}(\bm{\mathrm{J}}_2,\bm{\mathrm{J}}_1)
    Z_{bb}(\bm{\mathrm{J}}_2,\bm{\mathrm{J}}_1)
    }} \Big] \\
    &=& \frac12 \Big[ \log \left< \phi_{a|b}(0) \phi_{a|b}(L) \right>_{\bm{\mathrm{J}}_1,\bm{\mathrm{J}}_2} \nonumber \\
    && + \log \left< \phi_{a|b}(0) \phi_{a|b}(L) \right>_{\bm{\mathrm{J}}_2,\bm{\mathrm{J}}_1} \Big] \,,
\end{eqnarray}
where we have used $Z_{aa}(\bm{\mathrm{J}}_2,\bm{\mathrm{J}}_1) = Z_{aa}(\bm{\mathrm{J}}_1,\bm{\mathrm{J}}_2)$. 
Since $\bm{\mathrm{J}}_1$, $\bm{\mathrm{J}}_2$ are independent disorder configurations, after disorder average, we arrive at 
\begin{eqnarray}
    \overline{\log Q_{ab}} = \overline{\log \left< \phi_{a|b}(0) \phi_{a|b}(L) \right>} \,.
\end{eqnarray}

Next we show that while the expression involves the average outside the logarithm, it reveals universal conformal data of the b.c.c.\ operator. 
In fact, in disordered systems, the scaling dimension of the b.c.c.\ operator $\phi_{a|b}$ depends on the procedure of disorder average. 
In the scaling limit, the conformal symmetry leads to  
\begin{equation}
    \overline{\left< \phi_{a|b}(0)\phi_{a|b}(L) \right>^n} \simeq L^{-2h_{ab}(n)}\,,
\end{equation} 
where $h_{ab}(n)$ denotes the scaling dimension. 
Now, noting $\log \left< \phi_{a|b}(0)\phi_{a|b}(L) \right> = \partial_n \left< \phi_{a|b}(0)\phi_{a|b}(L) \right>^n|_{n=0}$, we arrive at
\begin{eqnarray}
\begin{split}
    \overline{\log Q_{ab}} &= \partial_n \overline{\left< \phi_{a|b}(0)\phi_{a|b}(L) \right>^n} \big|_{n=0} \\
    &= \partial_n  L^{-2h_{ab}(n)}|_{n=0} \\
    &\simeq -2 h_{ab}'(0) \log L\,.
\end{split}    
\end{eqnarray}
Hence, we show that the scaling dimension appearing in  Eq.~\eqref{eq:bcc} in the main text is universal data from the underlying BCFT, $\Delta_{ab} = h_{ab}'(0)$.

\paragraph{Boundary decoding threshold.}\ssll
In this section, we analyze the boundary threshold in the surface code. 
We compute the coherent information at $p_{\rm bulk} = 0$, which gives the boundary threshold $p_{\rm bdy,c}=\frac12$.  
For $p_{\rm bulk} = 0$, the spin can be set $\sigma_i = +1$, the partition function is determined by,
\begin{equation}
    Z(\{\eta_i\})=e^{-H_{\rm bdy}} \,, \quad H_{\rm bdy}=\sum_{i\in{\rm bdy}}J_i \equiv \sum_{i\in{\rm bdy}}J \eta_i \, .
\end{equation}
The coherent information is then
\begin{eqnarray}
\label{eq:CI of repetition code}
    && \mathcal{I}_{\rm rep}(p_{\rm bdy})=1+\sum_{\eta_i=\pm}\frac{Z(\{\eta_i\})\log_2{\left[\frac{Z(\{\eta_i\})}{Z(\{\eta_i\})+Z(\{-\eta_i\})}\right]}}{\sum_{\eta'_i=\pm}Z(\{\eta'_i\})} \nonumber \\
    &&=1-\sum_{i=0}^N {N \choose i} p_{\rm bdy}^{\,i}
    (1-p_{\rm bdy})^{N-i} \nonumber \\
    &&\times\log_2{\left[1+\left(\frac{1-p_{\rm bdy}}{p_{\rm bdy}}\right)^{(2i-N)}\right]}\\
    &&\approx 1-\int_0^1 {\rm d}x P(x) \log_2{\left[1+\left(\frac{1-p_{\rm bdy}}{p_{\rm bdy}}\right)^{N(2x-1)}\right]} \,,\nonumber 
\end{eqnarray}
where $x=i/N$, $N$ is the number of boundary spins, and
\begin{equation}
\begin{split}
    P(x)=&\
    \sqrt{\frac{N}{2\pi x(1-x)}}
    e^{-N D(x||p_{\rm bdy})} \,,
    \\
    D(x||p)=&\ x\log\frac{x}{p}
    +(1-x)\log\frac{1-x}{1-p} \,.
\end{split}
\end{equation}
Here $P(x)$ follows from Stirling's approximation, and $D(x||p)\geq0$ with equality holding only at $x=p$.

We now focus on $p_{\rm bdy}< \frac12$. 
For $x<\frac12$, the logarithmic factor in Eq.~\eqref{eq:CI of repetition code} is exponentially small in $N$. 
For $x>\frac12$, the logarithmic factor grows at most linearly with $N$, while $P(x)$ is exponentially suppressed because $D(x||p_{\rm bdy})>0$. 
Thus the dominant contribution comes from the vicinity of $x=\frac12$. 
With $D(\frac12||p_{\rm bdy})=-\frac{1}{2}\log\left[4p_{\rm bdy}(1-p_{\rm bdy})\right]$, we obtain the asymptotic form
\begin{equation}
\label{eq:repetition code F CI app}
    \mathcal{I}_{\rm rep}(p_{\rm bdy})=1-A(p_{\rm bdy})N^{-1/2}\left[4p_{\rm bdy}(1-p_{\rm bdy})\right]^{N/2}+\cdots \, .
\end{equation}
Therefore, for any fixed $p_{\rm bdy}<\frac12$, $\mathcal{I}_{\rm rep}\rightarrow1$ as $N\rightarrow\infty$. 
At $p_{\rm bdy}=\frac12$, the two logical sectors have equal weight, and one obtains $\mathcal{I}_{\rm rep}=0$.
This shows that the boundary threshold is given by $p_{\rm bdy,c} =\frac12$.

\newpage
\onecolumngrid

\renewcommand{\theequation}{S\arabic{equation}}
\renewcommand{\thefigure}{S\arabic{figure}}
\renewcommand\figurename{Supplementary Figure}
\renewcommand\tablename{Supplementary Table}
\newpage

\section*{Supplemental Material}

\section{I. Details of tensor-network calculation}

\begin{figure}[b]
    \centering
    \includegraphics[width=0.5\linewidth]{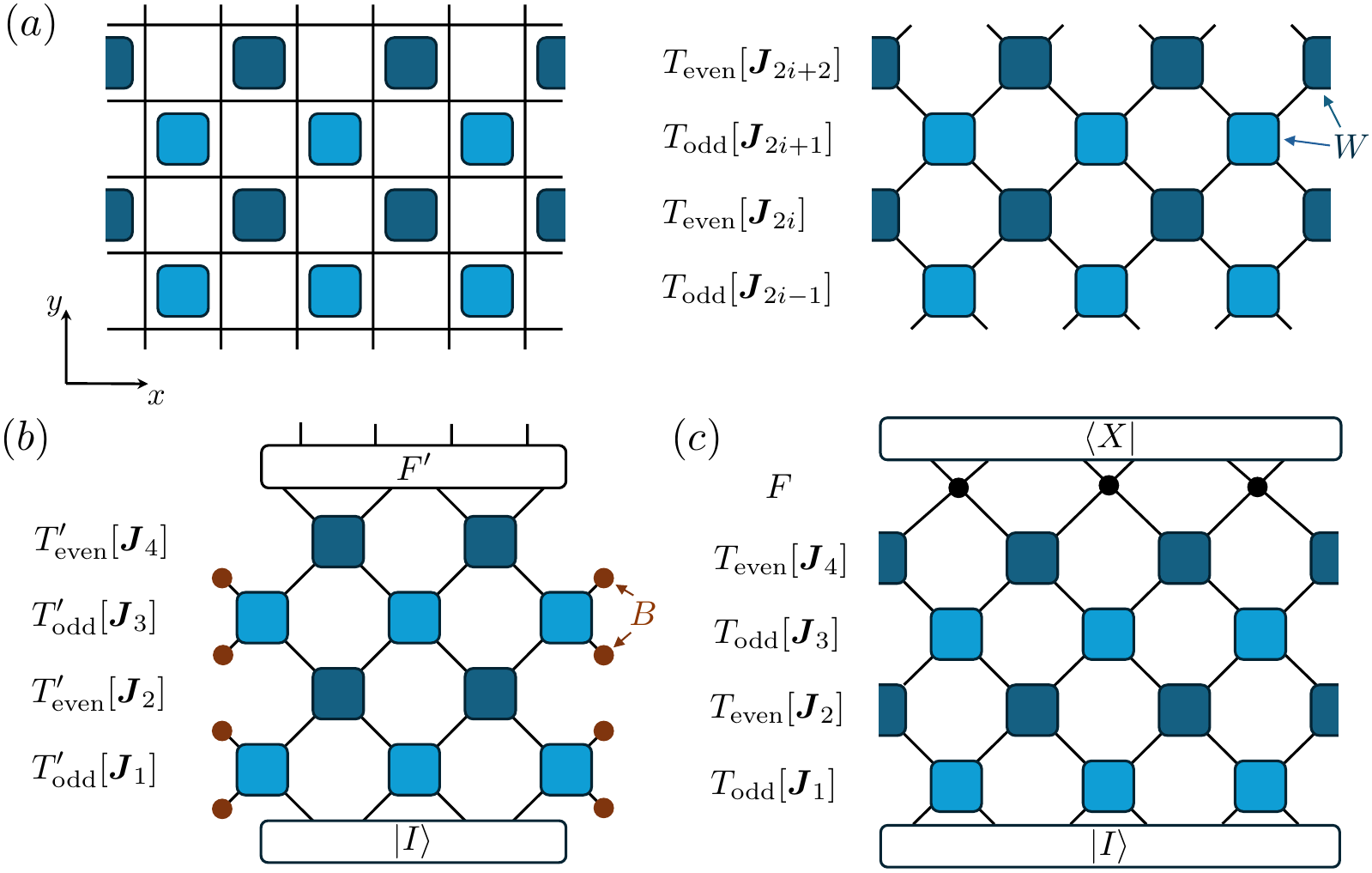}
    \caption{Illustration of the tensor-network method. 
    (a) The partition function is represented by a contraction of the tensor network, where each small square indicates a local tensor $W$ encoding the local Boltzmann weight contributed from the surrounding four bonds. The transfer matrix $T_\text{odd}[\boldsymbol{J}_{2i-1}]$ ($T_\text{even}[\boldsymbol{J}_{2i}]$) is composed by a layer of $W$ tensors colored by light (dark) blue.
    (b) The tensor network with fixed boundary spin orientations in the open string channel, represents the calculation of Eq.~\eqref{eq:psi_theta}. The boundary condition is imposed by the boundary $B$ tensors. 
    (c) The tensor network with a periodic boundary condition in the closed string channel, represents the denominator $\left< X | \Psi \right>$ in Eq.~(7) of the main text. The numerator can be obtained by applying $\sigma_{i}^{z}\sigma_{j}^{z}$ on $\left| \Psi \right>$ before its contraction with $\left| X \right>$.}
    \label{fig:tn}
\end{figure}

In this section, we provide the technical details of the tensor-network construction used in the main text. 
We first discuss the tensor-network representation of the random-bond Ising model (RBIM) at the Nishimori multicritical point, then introduce the boundary tensors that implement different microscopic boundary spin orientations. 
We also explain the role of the final-layer transfer matrix, which is required to obtain the correct path-integral interpretation of wavefunction overlaps and von Neumann entropies. 
Finally, we describe how the boundary entropy and boundary-condition-changing (b.c.c.) data are extracted from the resulting tensor-network states, and present the detail of the matrix-product-state (MPS) simulation. 

\subsection{A. Tensor-network representation}

The partition function of RBIM can be represented by a tensor network as illustrated in Fig.~\ref{fig:tn}(a).
We define our building block $W_{\sigma_1,\sigma_2}^{\tilde \sigma_1, \tilde \sigma_2 } = e^{\sum_{\langle \sigma,\sigma' \rangle} J_{\sigma \sigma'} \sigma \sigma'} $ with $\sigma, \sigma' \in \{\sigma_1,\sigma_2, \tilde \sigma_1, \tilde \sigma_2 \}$, and $J_{\sigma \sigma'}$ is the corresponding random bond coupling between spins $\sigma$ and $\sigma'$. It encodes the local Boltzmann weight contributed from the surrounding four bonds.
To be concrete, we consider a $L_x \times L_y$ lattice with periodic boundary conditions and take $L_x$ and $L_y$ both to be even. 
The transfer matrix can be obtained from the $W$ tensor as
\begin{equation} \label{eq:TM_bulk}
\begin{split}
    T_{\rm odd}[\bm J] = \sum_{\bm \sigma,\tilde{\bm \sigma}} \prod_{i=1}^{L_x/2} W_{\sigma_{2i-1},\sigma_{2i}}^{\tilde\sigma_{2i-1},\tilde\sigma_{2i}} \left|  \tilde{\bm \sigma} \right> \left< \bm \sigma \right| \,, \qquad
    T_{\rm even}[\bm J] = \sum_{\bm \sigma,\tilde{\bm \sigma}} \prod_{i=1}^{L_x/2}W_{\sigma_{2i},\sigma_{2i+1}}^{\tilde\sigma_{2i},\tilde\sigma_{2i+1}} \left|  \tilde{\bm \sigma} \right> \left< \bm \sigma \right| \,,
\end{split}
\end{equation}
where $\bm \sigma = \{\sigma_1, ..., \sigma_{L_x}\}$, $\tilde{\bm \sigma}= \{\tilde \sigma_1, ..., \tilde\sigma_{L_x}\}$ and the random variables contained in $W$ tensors are abbreviated as $\bm J$.  
We have $\sigma_{L_x + 1} = \sigma_1$ and $\tilde\sigma_{L_x + 1} = \tilde\sigma_1$ in $T_{\rm even}$ for the periodic boundary condition. 
Then the partition function with the periodic boundary condition is 
\begin{equation}
    Z = \Tr\left[ \prod_{j=1}^{L_y/2} T_{\rm even}[\bm J_{2j} ] T_{\rm odd}[\bm J_{2j-1} ] \right]\,.
\end{equation}

\subsection{B. Boundary tensors, open-string transfer matrix, and final-layer transfer matrix}

As mentioned in the main text, fixed boundary spin orientations can be implemented by the boundary tensor, $B_\sigma^\theta = \cos(\frac{\pi}4 - \frac\theta2) \delta_{\sigma,+1} + \sin(\frac{\pi}4 - \frac\theta2) \delta_{\sigma,-1}$, where $\theta$ characterizes the spin orientation,
\begin{equation}
    \left< B^\theta \right| \vec \sigma \left| B^\theta \right> = \cos \theta \hat i + \sin \theta \hat k\,,
\end{equation}
where $\left| B^\theta \right> = \sum_{\sigma} B_\sigma^\theta \left| \sigma \right>$ and $\vec \sigma = \sigma^x\hat{i}+\sigma^y\hat{j} +\sigma^z\hat{k}$ is the Pauli matrix.
In particular, $\theta = 0$ corresponds to the free boundary condition and $\theta = \pm \frac\pi2$ to fixed boundary conditions $\sigma = \pm 1$. 
To incorporate this boundary condition, we define the transfer matrix
\begin{equation} 
\label{eq:TM_bdy}
    T'_{\rm odd}[\bm J] = \sum_{\bm \sigma,\tilde{\bm \sigma}} B_{\tilde \sigma_1}^\theta B_{\sigma_1}^\theta \left(\prod_{i=1}^{L_x/2} W_{\sigma_{2i-1},\sigma_{2i}}^{\tilde\sigma_{2i-1},\tilde\sigma_{2i}} \right) B_{\tilde \sigma_{L_x}}^\theta B_{\sigma_{L_x}}^\theta \left|  \tilde{\bm \sigma}_{b} \right> \left< \bm \sigma_{b} \right| \,, \qquad
    T'_{\rm even}[\bm J] = \sum_{\bm \sigma_b,\tilde{\bm \sigma}_b} \prod_{i=1}^{L_x/2-1}W_{\sigma_{2i},\sigma_{2i+1}}^{\tilde\sigma_{2i},\tilde\sigma_{2i+1}} \left|  \tilde{\bm \sigma}_{b} \right> \left< \bm \sigma_{b} \right| \,,
\end{equation}
where $\bm \sigma_b = \{\sigma_2, ..., \sigma_{L_x-1}\}$, $\tilde{\bm \sigma}_b= \{\tilde \sigma_2, ..., \tilde\sigma_{L_x-1}\}$ denotes the spin variables in the bulk, while the boundary spins have fixed orientation. 

We use time-evolving block
decimation (TEBD) to obtain the conformal data for different boundary spin orientations. 
As shown in Fig.~\ref{fig:tn}(b), we consider an initial state $\left|I\right>$ (the detail of which is not important) and evolve according to the transfer matrix for sufficiently large steps (equivalently, $L_y$). 
To guarantee convergence, we have set $L_{y} = 10 L_{x}$, ensuring that the results have reached their steady-state value. 
For simplicity, $L_{x}$ is also denoted as $L$ in the following discussion.
After obtaining the steady state, a special last-layer transfer matrix $F'$ is applied, namely, 
\begin{equation}
    \label{eq:psi_theta}
    \left| \Psi_\theta(\bm{\mathrm{J}}) \right> = \frac1{\mathcal N'} F' \prod_{j=1}^{L_y/2} T'_{\rm even}[\bm J_{2j} ] T'_{\rm odd}[\bm J_{2j-1} ] \left|I \right> \,, 
\end{equation}
where $\mathcal N'$ is the normalization factor for the wavefunction and $F' = \sum_{\bm \sigma_b} e^{- \sum_{i=2}^{L_x-2} (-1)^i J_{\sigma_i\sigma_{i+1}} \sigma_i \sigma_{i+1}/2}  \left|  {\bm \sigma}_b \right> \left< \bm \sigma_b \right|$ in the final layer. 
This additional layer, $F'$, is introduced to ensure the correct evaluation of the von Neumann entropy and the corresponding wavefunction overlaps. 
The key requirement for this is that the state $\left| \Psi_\theta (\bm{\mathrm{J}})\right>$ produced by the tensor-network evolution must admit a clean path-integral interpretation, such that 
\begin{enumerate}
    \item the overlap $\left< \Psi_{\theta'} (\bm{\mathrm{J}}_2)| \Psi_{\theta}(\bm{\mathrm{J}}_1) \right>$ is proportional to the classical partition function on a cylinder with boundary orientations $\theta'$ and $\theta$;
    \item the reduced density matrix $\rho_A(\bm{\mathrm{J}}) = \Tr_B [\left| \Psi_\theta(\bm{\mathrm{J}}) \right> \left< \Psi_\theta(\bm{\mathrm{J}}) \right| ]$ generates the same replica path integral as the standard construction of von Neumann entropy from twist operators in a 2D statistical mechanics model. 
\end{enumerate}
Both conditions are sensitive to how the last layer of tensors is chosen in the brick-wall decomposition.
The introduction of the special final-layer transfer matrix $F'$ is precisely to ensure that the resulting state $\left| \Psi_{\theta}(\bm{\mathrm{J}}) \right>$ satisfies these requirements, and furthermore, this $F'$ is uniquely determined by these requirements in our case.

\subsection{C. Wavefunction overlap and von Neumann entropy}

In the open-string channel, we alternate the transfer matrices $T_{\rm odd}'$ and $T_{\rm even}'$ constructed from the local tensor $W$ and the boundary tensor $B$ [see Fig.~\ref{fig:tn}(b)].
In the brick-wall geometry, however, the application of $T_{\rm even}' T_{\rm odd}'$ does not cover all bonds of the corresponding two rows of the original lattice: one staggered set of bonds in the final time slice is left out. 
If we were to ignore these bonds, then the resulting geometry is not the original RBIM cylinder but one with a microscopic defect line.

To avoid this, we insert an additional transfer matrix $F'$ that  precisely corrects the Boltzmann weights along the ``seam'' where bra and ket are sewn together, as illustrated in the left panel in Fig.~6 of the End Matter.
With this choice, the overlap becomes
\begin{eqnarray} \label{eq:overlap_Z}
    \left< \Psi_{\theta'}(\bm{\mathrm{J}}_2) | \Psi_\theta (\bm{\mathrm{J}}_1)\right> \propto \sum_{\{{ \sigma }\}} e^{\sum_{\left< ij\right>}  J_{ij} \sigma_i \sigma_j} B_{\theta}(\sigma_{\rm bdy}^{\rm(bottom)})B_{\theta'}(\sigma^{\rm(top)}_{\rm bdy}) 
    \equiv Z_{\theta',\theta}(\bm{\mathrm{J}}_2,\bm{\mathrm{J}}_1)\,,    
\end{eqnarray}
where $B_{\theta}$ and $B_{\theta'}$ implement the boundary conditions $\theta$ and $\theta'$, respectively. 
More precisely, $B_{\theta}(\sigma_{\rm bdy}^{\rm(bottom)}) = \prod_{\sigma \in {\rm bdy, bottom}} B^\theta_{\sigma} $, and $B_{\theta'}(\sigma^{\rm(top)}_{\rm bdy}) = \prod_{\sigma \in {\rm bdy, top}} B^{\theta'}_{\sigma}$, where ``bdy'' indicates the boundary sites, and bottom and top boundary sites are illustrated in the middle panel of Fig.~6 by the brown and orange dots, respectively. 
$\bm{\mathrm{J}}_1$ and $\bm{\mathrm{J}}_2$ are two independent random bond configurations except for the last layer $F'$. 
The reason we use independent bond configurations is to generate a full random ensemble for the 2D system, as $\bm{\mathrm{J}}_1$ and $\bm{\mathrm{J}}_2$ denote random bond configurations for the bottom and top half-plane, respectively, in the middle panel. 
Then we calculate Eq.~(6) of the main text to obtain the typical scaling dimension for the b.c.c. operators.

Regarding the calculation for von Neumman entropy, consider two subsystems, $A=\{1,...,L_A\}$ and $B= \{L_A+1,...,L\}$, the von Neumann entropy of the subsystem $A$ is defined as $S_{A} [\Psi_\theta(\bm{\mathrm{J}})] = - \Tr[\rho_A(\bm{\mathrm{J}}) \log \rho_A(\bm{\mathrm{J}}) ]$, where the reduced density matrix $\rho_A$ is obtained by tracing out subsystem $B$, $\rho_A(\bm{\mathrm{J}}) = \Tr_{B} [ \left|\Psi_\theta(\bm{\mathrm{J}}) \right> \left<\Psi_\theta(\bm{\mathrm{J}}) \right| ]$. 
While the RBIM is a classical model, we show in the End Matter that the half-chain (i.e., $L_A = L/2$) von Neumann entropy gives 
\begin{equation} \label{seq:vN_entropy}
\begin{split}
    \overline{S_{A}[\Psi_\theta]} \equiv  \sum_{\bm{\mathrm{J}}} P(\bm{\mathrm{J}}) S_{A}[\Psi_\theta(\bm{\mathrm{J}})]
     = \frac{c_{\rm vN}}6  \log \frac{L}\pi + \tilde S_{\rm bdy}(\theta) + s_{\rm UV}   \,.
\end{split}
\end{equation}
where $\overline{(\cdot)}$ is the average of the random bond configuration $\bm{\mathrm{J}}$.
If the boundary tensor is chosen randomly, the disorder average $\overline{(\cdot)}$ is understood to include the average over the boundary-tensor ensemble as well.

\subsection{D. Details of the matrix-product-state simulation}

In our tensor-network calculations, the state of the system during the transfer-matrix evolution is represented as an MPS. 
To keep the computation tractable as the transfer matrix is applied layer by layer, we performed singular value decompositions (SVDs) after applying local tensors to update the MPS to control the bond dimension $\chi$ of the MPS. 
Let $\{ \lambda_{i} \}$ be the spectrum of singular values from the SVD across a given MPS bond, which is sorted in descending order and normalized such that $\sum_{i} \lambda_{i}^{2} = 1$. 
The truncation error $\epsilon$ is defined as the weight that we discard, $\epsilon \equiv \sum_{i > \chi} \lambda_{i}^{2} = 1 - \sum_{i \le \chi} \lambda_{i}^{2}$, where $\chi$ is the number of retained singular values (i.e., the MPS bond dimension).
To ensure a high numerical accuracy, we utilize a dynamically adaptive bond dimension $\chi$ during the evolution. 
Rather than fixing a hard maximum bond dimension, we set a cutoff threshold $10^{-10}$. 
This means that at each evolution step and across every MPS bond, we retain as many singular values as necessary to ensure that the discarded weight $\epsilon$ never exceeds $10^{-10}$.

\section{II. Benchmark of the tensor-network wavefunction method on the clean Ising model}

To demonstrate the validity and workflow of our numerical procedure before applying it to the disordered system, we benchmark our tensor-network method on the well-understood two-dimensional classical Ising model. 
This clean limit serves as a pedagogical example to illustrate how the boundary entropy, the boundary renormalization-group flow, and the scaling dimensions of the b.c.c. operators can be accurately extracted.

The clean Ising model on a square lattice undergoes a continuous phase transition at the critical coupling strength $J_\text{c} = \frac{\ln(1+\sqrt{2})}{2}$. 
At this critical point, its bulk physics is described by a CFT with central charge $c = 1/2$. 
On the boundary, by introducing a tuning parameter $\theta \in [0, \pi/2]$ analogous to the main text, we can interpolate between two different conformal boundary conditions: the free boundary condition (denoted by $f$ at $\theta = 0$) and the fixed boundary condition (denoted by $\pm$ at $\theta = \pm \pi/2$). 

\begin{figure}
    \centering
    \includegraphics[width=0.6\linewidth]{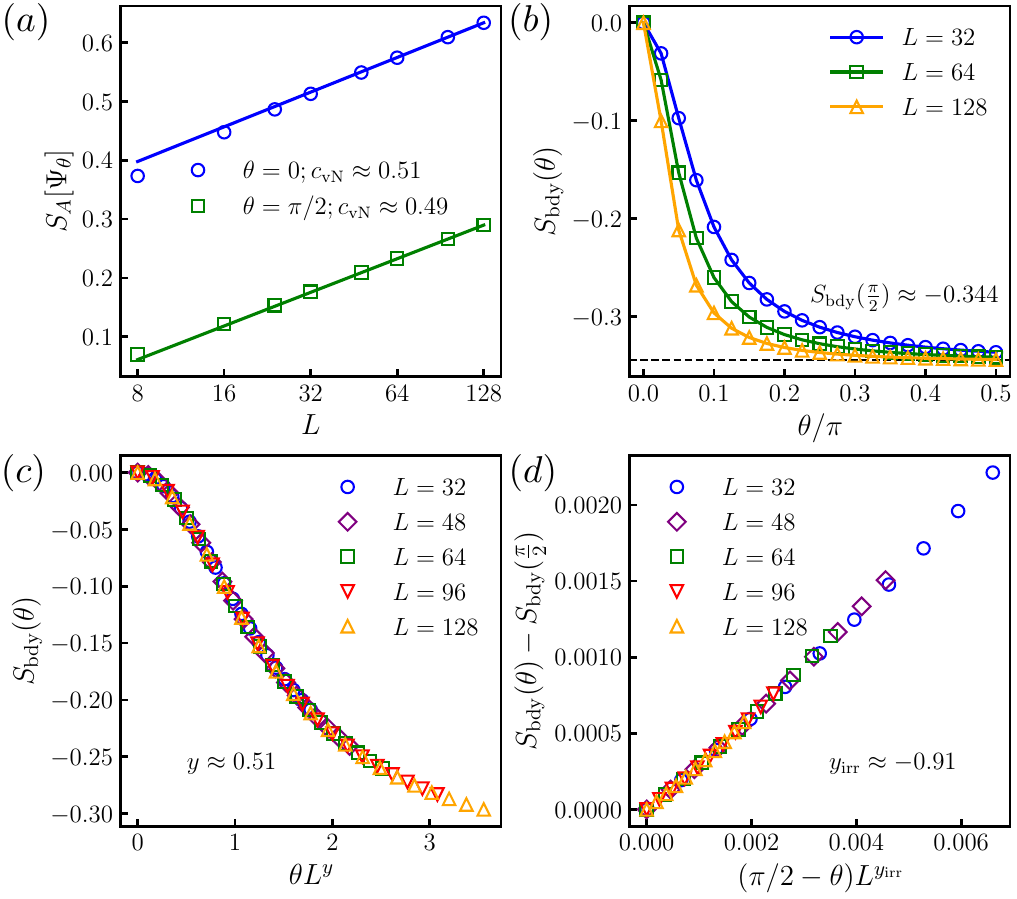}
    \caption{Boundary entropy and boundary renormalization-group flow for the two-dimensional clean Ising model. (a) Half-chain von Neumann entropy $S_{A}[\Psi_{\theta}]$ versus the size $L$ for $\theta = 0$ and $\pi/2$. The logarithmic-law fittings estimate the central charge $c_{\rm vN} \approx 0.51$ and $0.49$ for $\theta = 0$ and $\pi/2$, respectively.
    (b) The relative boundary entropy $S_{\rm bdy}(\theta)$ versus $\theta$ for different $L$. The value at $\theta = \pi/2$ is estimated to be $-0.344$\,.
    (c) Data collapse near $\theta_\ast = 0$ with $y \approx 0.51$\,. (d) Data collapse near $\theta_\ast = \pi/2$ with $y_{\rm irr} \approx -0.91$\,.}
    \label{fig:entropy_clean}
\end{figure}

We first compute the half-chain von Neumann entanglement entropy $S_{A}[\Psi_{\theta}]$ of the critical clean Ising model under different boundary conditions. 
As shown in Fig.~\ref{fig:entropy_clean}(a), the entanglement entropy scales logarithmically with the system size $L$, $S_{A}[\Psi_{\theta}] = \frac{c_\text{vN}}{6} \log{\frac{L}{\pi}} + \tilde{S}_\text{bdy}(\theta) + s_\text{UV}$. 
By fitting the data, we extract the central charge $c \approx 0.51$ for $\theta = 0$ and $c \approx 0.49$ for $\theta = \pi/2$, both in agreement with the theoretical value.

More importantly, the intercept of the scaling relation provides the boundary entropy. 
For the clean Ising CFT, the theoretical values are, respectively, $0$ and $-\frac{1}{2}\log2 \approx -0.347$ for the free and fixed boundary conditions. 
As shown in Fig.~\ref{fig:entropy_clean}(b), our numerical result yields $S_\text{bdy}(\frac{\pi}{2}) \equiv \tilde{S}_\text{bdy}(\frac{\pi}{2}) - \tilde{S}_\text{bdy}(0) \approx -0.344$, which perfectly matches the theoretical value. 

By tuning the angle $\theta$ away from $0$, we effectively introduce a boundary magnetic field that drives a boundary RG flow towards the fixed boundary condition. 
To study this boundary RG flow, we perform a finite-size scaling analysis on $S_\text{bdy}(\theta)$. 
As shown in Fig.~\ref{fig:entropy_clean}(c) and Fig.~\ref{fig:entropy_clean}(d), the relative boundary entropy calculated on different system sizes collapses onto a single curve as a function of the rescaled variable $\theta L^{y}$ [or $(\frac{\pi}{2}-\theta)L^{y_\text{irr}}$]. 
The optimal data collapses are achieved with $y \approx 0.51$ and $y_\text{irr} \approx -0.91$\,.

\begin{figure}
    \centering
    \includegraphics[width=0.6\linewidth]{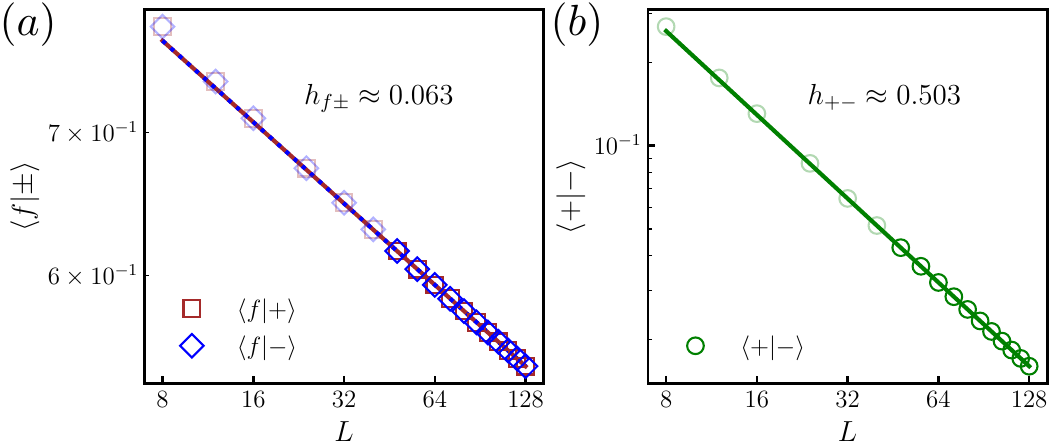}
    \caption{Wavefunction overlaps versus lattice size $L$ for the two-dimensional clean Ising model. The least-squares fittings according to $\langle a | b \rangle \sim L^{-2h_{ab}}$ estimate the scaling dimensions of the b.c.c operator, $h_{f\pm} \approx 0.063$ and $h_{+-} \approx 0.503$\,. The subscripts $f$ and $\pm$ denote the free and fixed $\pm$ boundary states, respectively. The six data points with the smallest values of $L$ were excluded from the fitting.}
    \label{fig:bcc_clean}
\end{figure}

Finally, we demonstrate the extraction of the scaling dimensions of the b.c.c. operators using the wavefunction overlap method. 
As shown in Fig.~\ref{fig:bcc_clean}, we compute the overlaps between wavefunctions for free ($\theta = 0$) and fixed $\pm$ ($\theta = \pm \frac{\pi}{2}$) boundary conditions. 
For the overlap between the free and fixed boundary conditions, the theoretical value of the scaling dimension is $h_{f\pm} = \frac{1}{16}$. 
Our numerical result yields $h_{f\pm} \approx 0.063$ [see Fig.~\ref{fig:bcc_clean}(a)]. 
For the overlap between the opposite fixed boundary conditions (i.e., $+$ and $-$), the b.c.c scaling dimension is $h_{+-} = \frac{1}{2}$. 
Our numerical result yields $h_{+-} \approx 0.503$ [see Fig.~\ref{fig:bcc_clean}(b)]. 
The remarkable agreement between our numerical results and the exact BCFT data firmly establishes the reliability of the wavefunction method, validating its subsequent applications to the disordered case, with the necessary modifications detailed in the main text and End Matter.

\section{III. Determination of the absolute boundary entropy for the free boundary condition}

In the main text, the boundary RG flow is analyzed by using the relative boundary entropy difference, taking the free boundary condition ($\theta = 0$) as a convenient reference.
In this  section, we explicitly extract the absolute boundary entropy of the free boundary condition $\tilde S_\text{bdy}(\theta=0)$ for both the clean Ising (as a benchmark) and the RBIM at the free boundary fixed point. 

Extracting the absolute boundary entropy $\tilde S_\text{bdy}(\theta)$ directly from the half-chain von Neumann entanglement entropy $S_{A}[\Psi_{\theta}]$ (we neglect the overline in $S_A[\Psi_\theta]$ for simplicity) is often challenging due to the presence of the  non-universal constant $ s_\text{UV}$ . 
To eliminate this constant, we employ a subtraction scheme using the entanglement entropy $S_{A}[\Psi_\text{PBC}]$ of the system under periodic boundary conditions (PBC)~\cite{igloi2008jsm,xavier2020prb}.
The half-chain entanglement entropy for a system of size $L$ under PBC and for a system with open boundaries (described by $\theta$ here) scales as follows~\cite{calabrese2004entanglement,igloi2008jsm}:
\begin{align}
  S_{A}[\Psi_\theta] & = \frac{c_\text{vN}}{6} \log \frac{2L}{\pi} +  s_\text{UV} + \tilde S_\text{bdy}(\theta) \, , \\
  S_{A}[\Psi_\text{PBC}] & = \frac{c_\text{vN}}{3} \log \frac{L}{\pi} + 2 s_\text{UV} \, .
\end{align}
By subtracting half of the entropy under PBC from the one with boundary condition $\theta$, the $s_\text{UV}$ term cancels out exactly. 
The absolute  boundary entropy can then be obtained by:
\begin{equation}
  \tilde S_\text{bdy}(\theta) = S_{A}[\Psi_{\theta}] - \frac{1}{2} S_{A}[\Psi_\text{PBC}] - \frac{c_\text{vN}}{6} \log 2 \, .
\end{equation}

\begin{figure}
    \centering
    \includegraphics[width=0.6\linewidth]{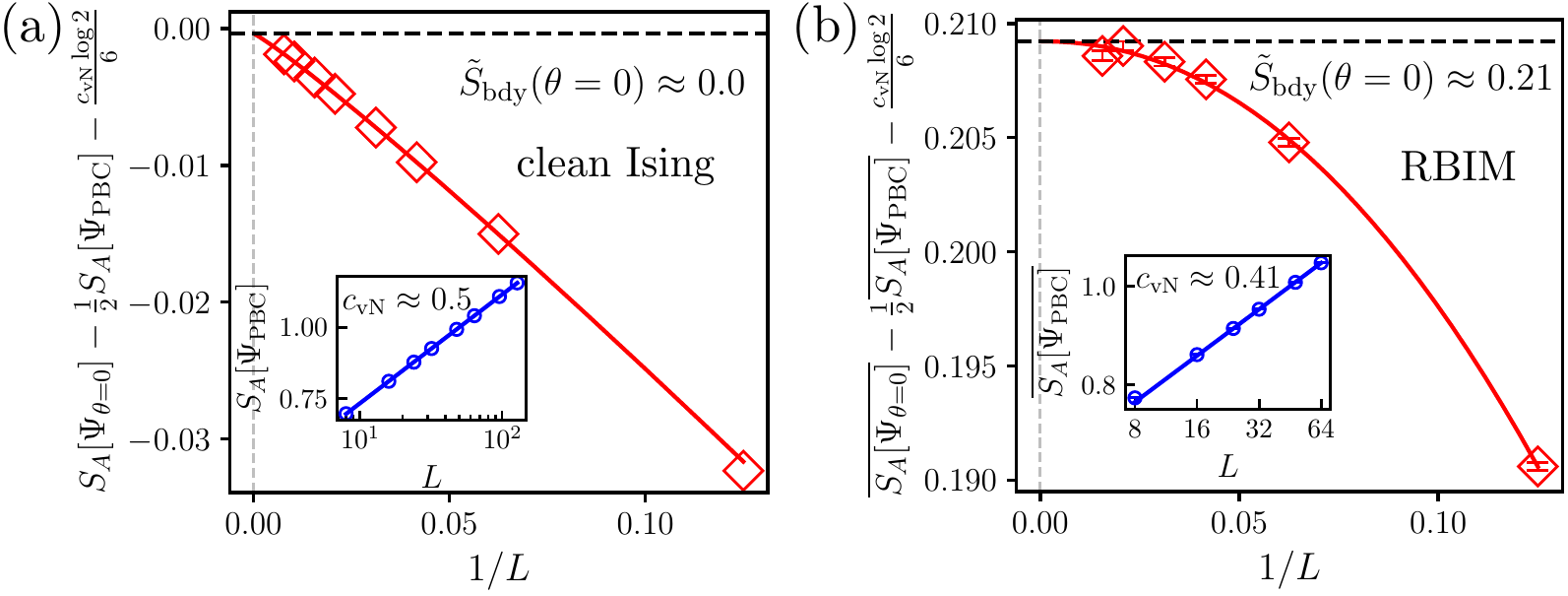}
    \caption{(a) The subtracted half-chain entanglement entropy, $S_{A}[\Psi_{\theta=0}] - \frac{1}{2} S_{A}[\Psi_\text{PBC}] - \frac{c_\text{vN}}{6}\log2$, versus $1/L$ for the clean Ising model ($c_\text{vN} = 0.5$ is used). The inset shows the logarithmic-law behavior of $S_{A}[\Psi_\text{PBC}]$ yielding $c_\text{vN} \approx 0.5$\,. The extrapolation according to $A \cdot L^{-d} + \tilde S_\text{bdy}(0)$ yields $\tilde S_\text{bdy}(0) \approx 0.0$\,. (b) The same subtraction protocol applied to the Nishimori multicritical point under free boundary condition ($\theta = 0$). The inset shows the logarithmic-law behavior of $S_{A}[\Psi_\text{PBC}]$ yielding $c_\text{vN} \approx 0.41$\,. The same extrapolation to the thermodynamic limit gives a nontrivial value for the boundary entropy $\tilde S_\text{bdy}(0) \approx 0.21$ ($c_\text{vN} = 0.41956$ is used~\cite{Putz2025NishimoriCharge}). Solid lines denote fits to the numerical data.}
    \label{fig:entropy_base}
\end{figure}

To benchmark this method, we first apply it to the clean Ising model under the free boundary condition $(\theta = 0)$. 
For the 2D Ising CFT, the central charge is exactly $c_\text{vN} = \frac12$ and
the boundary entropy for the free boundary condition is known to be $\tilde S_\text{bdy}(0) = 0$. 
In Fig.~\ref{fig:entropy_base}(a), we plot the difference $S_{A}[\Psi_{\theta=0}] - \frac{1}{2} S_{A}[\Psi_\text{PBC}] - \frac{c_\text{vN}}{6}\log2$ as a function of the inverse system size $1/L$. 
Using the theoretical value $c_\text{vN} = \frac12$, we extrapolate the data to the thermodynamic limit;
the intercept yields $\tilde S_\text{bdy}(0) \approx 0.0$, which is in perfect agreement with the exact analytical value. 

Next, we apply the same procedure to the Nishimori multicritical point under the free boundary condition ($\theta = 0$). 
As shown in Fig.~\ref{fig:entropy_base}(b), we obtain a nontrivial absolute boundary entropy $\tilde S_\text{bdy}(0) \approx 0.21$\,.
This precise determination of the absolute boundary entropy provides an anchor point. 
Together with the results shown in the main text, we can straightforwardly deduce the absolute boundary entropy values for other conformal boundary conditions in the boundary RG diagram, namely, $\tilde S_\text{bdy}(\theta = \pi/2) \approx -0.364$ for the fixed boundary condition and $\tilde S_\text{bdy}(\theta_W = \pi/2) \approx -0.268$ for the random boundary condition.

\section{IV. Numerical evidence for conformal symmetry} 

In this section, we provide strong numerical evidence for the conformal symmetry of the BCFT presented in the main text.

\begin{figure}
    \centering
    \includegraphics[width=0.35\linewidth]{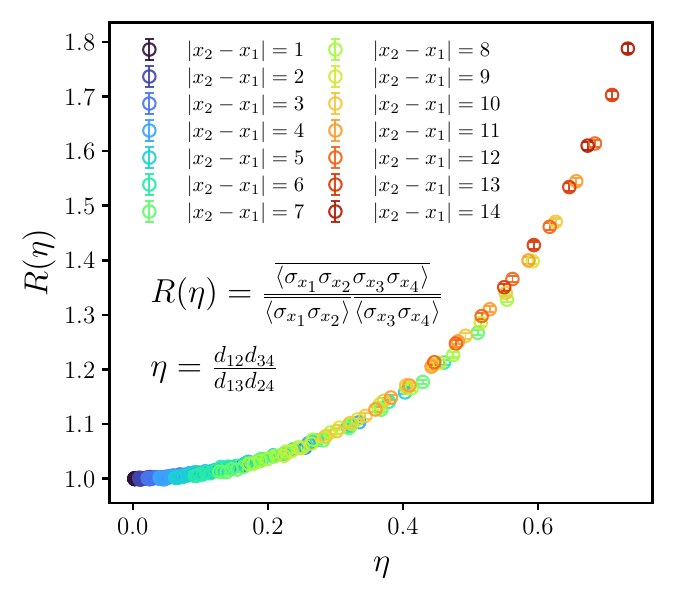}
    \caption{Data collapse of the normalized boundary four-point correlation function $R(\eta) \equiv \frac{\overline{\langle \sigma_{x_{1}} \sigma_{x_{2}} \sigma_{x_{3}} \sigma_{x_{4}} \rangle}}{\overline{\langle \sigma_{x_{1}} \sigma_{x_{2}} \rangle} \, \overline{\langle \sigma_{x_{3}} \sigma_{x_{4}} \rangle}}$ as a function of the conformal cross-ratio $\eta \equiv \frac{d_{12} d_{34}}{d_{13} d_{24}}$. Here, $d_{ij} = L/\pi \sin(\pi |x_i - x_j|/L)$ is the conformal chord length. The data are obtained under the free boundary condition for various spatial configurations of the four boundary spin positions with system size $L=64$. For simplicity, the positions are ordered as $x_{1} < x_{2} < x_{3} < x_{4}$ with the constraint $|x_{1} - x_{2}| = |x_{3} - x_{4}|$. For each value of $|x_{1} - x_{2}|$, we vary the distance $|x_{2} - x_{3}|$ to cover a wide range of the cross-ratio $\eta$.}
    \label{fig:c4_conformal}
\end{figure}

1. \emph{Four-point correlation functions and the conformal cross-ratio}---First, we computed the four-point correlation function of boundary spins under the free boundary condition. 
The disorder-averaged four-point correlation function is defined as $\overline{\langle \sigma_{x_{1}} \sigma_{x_{2}} \sigma_{x_{3}} \sigma_{x_{4}} \rangle}$, where $x_i$ denotes the position of the $i$-th boundary spin. 
Due to the presence of conformal invariance, this four-point correlation function should take the following form~\cite{francesco2012conformal}
\begin{equation}
    \overline{\langle \sigma_{x_{1}} \sigma_{x_{2}} \sigma_{x_{3}} \sigma_{x_{4}} \rangle} = \mathcal{F}(\eta_{1},\eta_{2}) \sum_{i<j} d_{ij}^{-2\Delta_{1}/3} \, ,
\end{equation}
where $\eta_{1} = \frac{d_{12}d_{34}}{d_{13}d_{24}}$, $\eta_{2} = \frac{d_{12}d_{34}}{d_{14}d_{23}}$, and $d_{ij} = L/\pi \sin(\pi |x_i - x_j|/L)$ is the conformal chord length with $\Delta_1$ the scaling dimension of the boundary spin operator. 
The function $\mathcal{F}(\eta_{1},\eta_{2})$ is an unknown universal function of the conformal cross-ratios $\eta_{1}$ and $\eta_{2}$, which should be independent of the system size $L$ or the specific positions of the four spins as long as the cross-ratios are fixed.
By using the relation $1/\eta_{1} = 1 + 1/\eta_{2}$, we can further simplify the four-point function to a single-variable function of $\eta \equiv \eta_{1}$ 
\begin{equation}
    \overline{\langle \sigma_{x_{1}} \sigma_{x_{2}} \sigma_{x_{3}} \sigma_{x_{4}} \rangle} = \frac{1}{\left(d_{12}d_{34}\right)^{2\Delta_{1}}} \left(\eta_{1}\eta_{2}\right)^{2\Delta_{1}/3} \mathcal{F}(\eta_{1},\eta_{2}) = \frac{1}{\left(d_{12}d_{34}\right)^{2\Delta_{1}}} \tilde{\mathcal{F}}(\eta) \, .
\end{equation}
This means that if we plot the normalized four-point correlation function $R(\eta) \equiv \frac{\overline{\langle \sigma_{x_{1}} \sigma_{x_{2}} \sigma_{x_{3}} \sigma_{x_{4}} \rangle}}{\overline{\langle \sigma_{x_{1}} \sigma_{x_{2}} \rangle} \, \overline{\langle \sigma_{x_{3}} \sigma_{x_{4}} \rangle}}$ as a function of the conformal cross-ratio $\eta$, all data points for different positions of the four spins should collapse onto a single universal curve.
As shown in Fig.~\ref{fig:c4_conformal}, we have evaluated various different spatial configurations of the four boundary spins. 
Strikingly, all data points collapse perfectly onto a single curve governed solely by the cross-ratio $\eta$.
This provides unambiguous evidence of the conformal invariance. 

\begin{figure}
    \centering
    \includegraphics[width=0.8\linewidth]{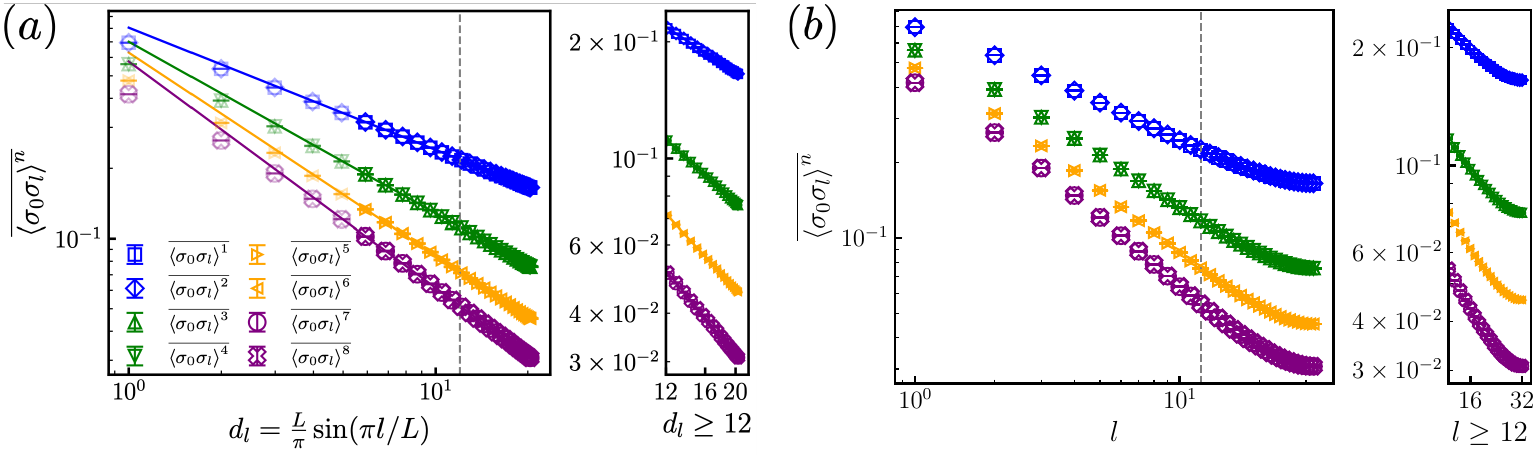}
    \caption{Log-log plots of the boundary spin-spin correlation moments, $\overline{\left< \sigma_0 \sigma_l \right>^n}$ ($n = 1$ to $8$), versus (a) the conformal chord length $d_{l} = L/\pi \sin(\pi l/L)$ and (b) the linear distance $l$ with size $L = 64$ for comparison. The narrow side panels zoom in on the long-distance tails ($d_{l} \ge 12$ and $l \ge 12$, respectively). These insets clearly demonstrate that using the conformal chord length perfectly maintains a pristine power-law scaling. In contrast, plotting against the linear distance $l$ leads to noticeable deviations at large distance.}
    \label{fig:correlation_sm}
\end{figure}

2. \emph{Two-point correlation functions governed by the conformal chord length}---While scale invariance predicts a power-law decay of the two-point boundary spin correlation function $\overline{\langle \sigma_{0} \sigma_{l} \rangle^{n}} \sim l^{-2\Delta_{n}}$, conformal invariance indicates that the correlation must scale with the conformally mapped chord length $d_{l} = L/\pi \sin(\pi l/L)$.
As shown in Fig.~\ref{fig:correlation_sm}, plotting the moments of the boundary spin-spin correlations against the linear distance $l$ leads to clear deviations at long distances. 
In striking contrast, plotting against the conformal chord length $d_{l}$ perfectly straightens the tail and maintains a power-law scaling up to the largest distance, providing further evidence of conformal invariance.

\begin{figure}
    \centering
    \includegraphics[width=0.6\linewidth]{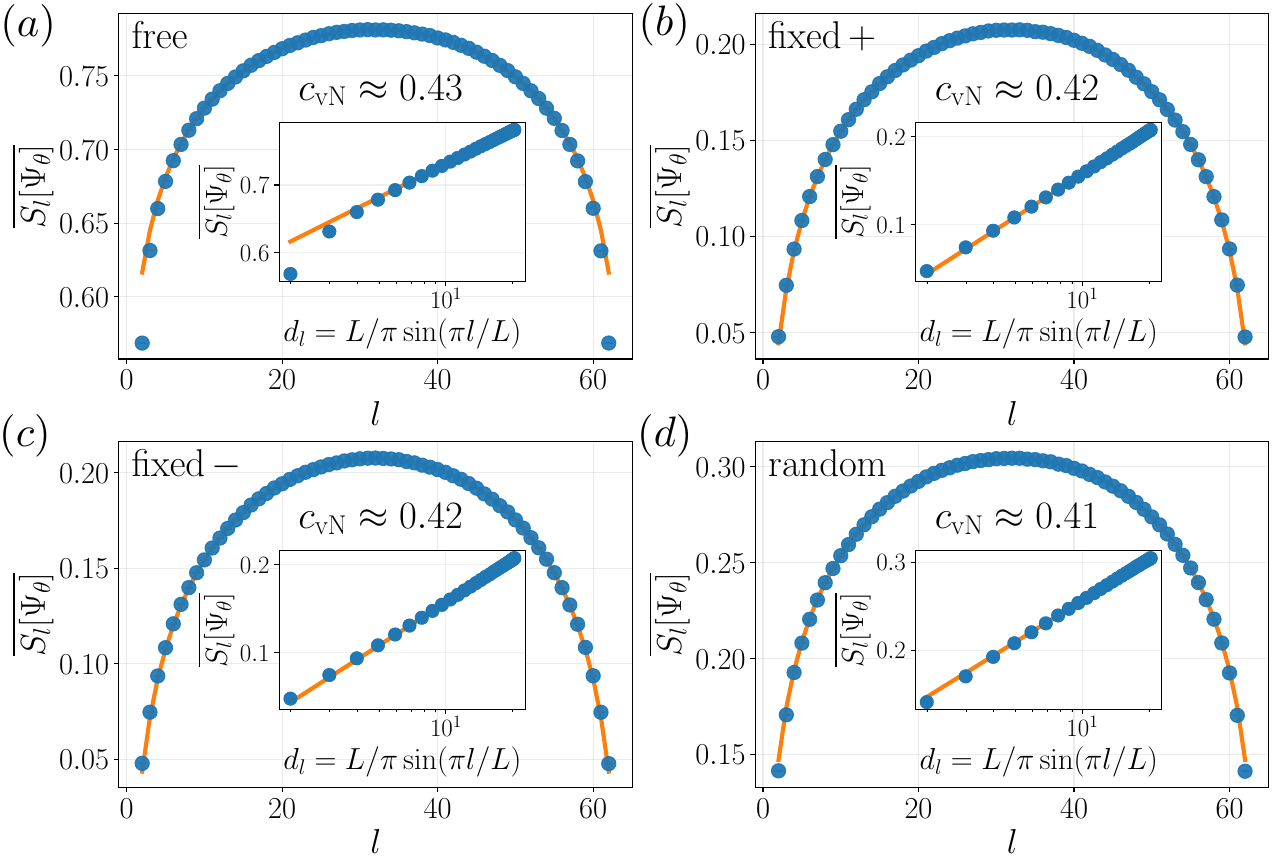}
    \caption{The bipartite von Neumann entropy $\overline{S_{l}[\Psi_{\theta}]}$ as a function of the subsystem size $l$ for the free (a), fixed $+$ (b), fixed $-$ (c), and random (d) boundary conditions with total size $L=64$. The inset in each panel shows the fit $\overline{S_{l}[\Psi_{\theta}]} = \frac{c_{\rm vN}}{6} \log d_{l} + {\rm const}$, where $d_{l} = L/\pi \sin(\pi l/L)$ is the conformal chord length. The excellent fits confirm the conformal invariance of the boundary fixed points.}
    \label{fig:ee_conformal}
\end{figure}

3. \emph{Scaling of bipartite von Neumann entropy with the conformal chord length}---Furthermore, we verified that the logarithmic scaling of the bipartite von Neumann entanglement entropy is also governed by the conformal chord length, precisely following the CFT prediction $\overline{S_{l}[\Psi_{\theta}]} = \frac{c_{\rm vN}}{6} \log d_{l} + {\rm const}$, where $d_{l} = L/\pi \sin(\pi l/L)$ is the conformal chord length.
As shown in Fig.~\ref{fig:ee_conformal}, this scaling form with the chord length has been verified not only for the free boundary condition but also for the fixed $\pm$ and random boundary conditions.

\section{V. Boundary RG calculation for the RBIM}

In this section, we summarize the perturbative derivation of boundary conformal data for the RBIM with Dirichlet boundary conditions near its upper critical dimension $d_c=6$. 
Following Refs.~\cite{le1988location,Le1989epsilon}, we consider an $m$-component classical spin $S(x)$ and introduce $n$ replicas. 
The bulk replicated effective action is
\begin{equation}
\label{eq:action_m_component_n_copy}
\begin{split}
    S_b=\int {\rm d}^d x\ &\left[\sum_\alpha\frac{1}{2}\left(r_m |M_\alpha(x)|^2+|\nabla M_\alpha(x)|^2\right)+\sum_{\alpha\beta}\frac{1}{2}\left(r_q |Q_{\alpha\beta}(x)|^2+|\nabla Q_{\alpha\beta}(x)|^2\right)\right.\\
    &\qquad\left.+2u \sum_{\alpha<\beta}M_\alpha(x) Q_{\alpha\beta}(x)M_\beta(x)+6w\sum_{\alpha<\beta<\gamma}Q_{\alpha\beta}(x)Q_{\beta\gamma}(x)Q_{\gamma\alpha}(x)\right],
\end{split}
\end{equation}
where $\alpha,\beta=1,\ldots,n$ are replica indices, and $M_\alpha(x)=\langle S_\alpha(x)\rangle$ and $Q_{\alpha\beta}(x)=\langle S_\alpha(x)S_\beta(x)\rangle(1-\delta_{\alpha\beta})$ are the ferromagnetic and spin-glass order parameters.
The RBIM corresponds to $m=1$. 
At the Nishimori fixed point, the theory has a gauge-like symmetry that exchanges $M_\alpha$ and $Q_{1\alpha}$ for $\alpha>1$, and this corresponds to $u=3w$. 
This condition is not imposed by hand: it is satisfied by the fixed point obtained from the bulk RG flow. 
In particular, in the replica limit $n\to0$ and for $m=1$, the fixed point is
\begin{equation}
    (u_*^2,w_*^2)=\left(\frac{\epsilon}{4K_d},\frac{\epsilon}{36K_d}\right) \,, \qquad \epsilon=6-d \,, \qquad K_d=\frac{1}{64\pi^3} \,.
\end{equation}
The bulk anomalous dimensions are $\eta_m=\eta_q=\frac{\epsilon}{3}$,
consistent with the Nishimori identity
$\overline{\langle\sigma_i\sigma_j\rangle^{2k-1}}=\overline{\langle\sigma_i\sigma_j\rangle^{2k}}$.

To study boundary criticality, we place the theory on a half-space $z\ge0$ and add the leading surface terms
\begin{equation}
\label{eq:surface_action}
    S_s=\int_{\partial\mathcal M}\left[\frac{c_m}{2}\sum_\alpha M_\alpha^2+\frac{c_q}{2}\sum_{\alpha<\beta}Q_{\alpha\beta}^2\right],
\end{equation}
where $c_m$ and $c_q$ tune the surface enhancement. 
The full action is $S=S_b+S_s$. 
We focus on the Dirichlet limit $c_m,c_q\to+\infty$~\cite{diehl1986field,diehl1996the}, corresponding to the ordinary surface universality class in the Landau description. 
This is the natural continuum candidate for the lattice free boundary condition studied in the main text. 
For Dirichlet boundary conditions, the leading boundary fields are the normal derivatives
\begin{equation}
    \widehat M_\alpha=\partial_z M_\alpha|_{z=0} \,,
    \qquad
    \widehat Q_{\alpha\beta}=\partial_z Q_{\alpha\beta}|_{z=0} \,.
\end{equation}
The critical mean-field Green's function in mixed momentum-position space is
\begin{equation}
\label{eq:mean field Greens function critical}
    G(\mathbf p,z,z')=\frac{1}{2p}\left(e^{-p|z-z'|} -e^{-p(z+z')}\right),
\end{equation}
where $p=|\mathbf p|$ is the momentum parallel to the boundary.

With the Dirichlet boundary condition, the action retains the gauge-like symmetry at the bulk fixed point satisfying $u=3w$.
This follows because, in the limit $c_m,c_q\rightarrow+\infty$, the boundary action in Eq.~\eqref{eq:surface_action} can be equivalently represented by the mean-field Green's function satisfying the Dirichlet boundary condition $G(\mathbf p,z=0,z')=0$ for both fields $M_\alpha$ and $Q_{\alpha\beta}$.
Combining this boundary Green's function with the bulk action, one can verify that the exchange $M_\alpha\leftrightarrow Q_{1\alpha}$ for $\alpha>1$ remains a symmetry.
As we show below, the boundary RG calculation is consistent with this gauge-like symmetry: the scaling dimensions of the boundary composite operators obey the multifractal relation $\Delta_{2k-1}=\Delta_{2k}$ for all $k\geq1$.

Now we compute the one-loop renormalization of the boundary fields $\widehat M_\alpha$ and $\widehat Q_{\alpha\beta}$.
We define the renormalized surface fields by $M_{s,0}=(Z_mZ_{m,s})^{1/2}M_{s,R}$ and $Q_{s,0}=(Z_qZ_{q,s})^{1/2}Q_{s,R}$, where $Z_m$ and $Z_q$ are the bulk field renormalization factors for $M_\alpha$ and $Q_{\alpha\beta}$, while $Z_{m,s}$ and $Z_{q,s}$ describe the additional renormalization of the corresponding surface fields $M_{\alpha,s}$ and $Q_{\alpha\beta,s}$.
The bulk factors $Z_m$ and $Z_q$ are obtained from the bulk RG analysis.
For the surface renormalization, summing the divergent parts of the relevant one-loop diagrams gives the boundary anomalous dimensions
\begin{equation}
\label{eq:anomalous dimension RG}
    \eta_{ms}=\frac{\partial\log Z_{ms}}{\partial\log\mu}=\frac{5(n-1)m u^2}{48\pi^3}\,,\qquad\eta_{qs}=\frac{\partial\log Z_{qs}}{\partial\log\mu}=\frac{5\left[u^2+9(n-2)m w^2\right]}{48\pi^3}\,.
\end{equation}
At the Nishimori fixed point and in the replica limit $n\rightarrow0,\ m=1$, these reduce to
\begin{equation}
    \eta_{ms}=\eta_{qs}=-\frac{5\epsilon}{3}\,.
\end{equation}
Therefore, the scaling dimensions of the leading boundary fields are
\begin{equation}
\label{eq:boundary_field_dimension}
    \Delta_1(d)=\Delta_2(d)=\frac{d-2}{2}+1+\frac{1}{2}\left(-\frac{\epsilon}{3}-\frac{5\epsilon}{3}\right)=\frac{d}{2}-\epsilon \, .
\end{equation}

More generally, we consider boundary composite operators associated with moments of the boundary spin correlation.
For the even moments $\overline{\langle S_s(x)S_s(0)\rangle^{2k}}$, the corresponding replica operator is
\begin{equation}
    O_s^{(2k)}=\prod_{i=1}^{k} Q_{\alpha_{2i-1}\alpha_{2i},s}+{\rm perm.},
\end{equation}
where ${\rm perm.}$ denotes the sum over all inequivalent pairings of the replica indices $\{\alpha_1,\ldots,\alpha_{2k}\}$.
This structure follows from two requirements.
First, the operator associated with the $2k$-th moment should be invariant under the permutation group $S_{2k}$ acting on the replica indices.
Second, among all operators with the same replica symmetry, the leading contribution comes from the operator with the lowest engineering dimension, which is constructed from the fundamental fields $M_{\alpha,s}$ and $Q_{\alpha\beta,s}$.
For an even moment, the lowest-dimensional $S_{2k}$-symmetric operator is therefore obtained by pairing the $2k$ replica indices into $k$ bilinears $Q_{\alpha\beta,s}$.
For example,
\begin{equation}
    O_s^{(4)}=Q_{\alpha_1\alpha_2,s}Q_{\alpha_3\alpha_4,s}+ Q_{\alpha_1\alpha_3,s}Q_{\alpha_2\alpha_4,s}+Q_{\alpha_1\alpha_4,s}Q_{\alpha_2\alpha_3,s}.
\end{equation}
Similarly, for the odd moments $\overline{\langle S_s(x)S_s(0)\rangle^{2k-1}}$, one replica index remains unpaired.
The leading boundary composite operator is then constructed from one boundary field $M_{\alpha,s}$ and $k-1$ boundary fields $Q_{\alpha\beta,s}$:
\begin{equation}
    O_s^{(2k-1)}=M_{\alpha_1,s}\prod_{i=1}^{k-1} Q_{\alpha_{2i}\alpha_{2i+1},s}+{\rm perm.}.
\end{equation}
Here ${\rm perm.}$ again denotes the sum over all inequivalent permutations of the replica indices, so that $O_s^{(2k-1)}$ transforms symmetrically under $S_{2k-1}$.

We now extract the anomalous dimension of $O_s^{(2k)}$ by evaluating the correlation function $\langle Q_{\alpha_1\alpha_2,s}\cdots Q_{\alpha_{2k-1}\alpha_{2k},s} O_s^{(2k)}\rangle$.
Summing the leading one-loop diagrams gives the anomalous contribution $\eta_{2k}=\frac{k(2-5k)}{3}\epsilon$.
Therefore, the scaling dimension of the even-moment boundary composite operator is
\begin{equation}
    \Delta_{2k}(d)=\frac{kd}{2}+\frac{k(2-5k)}{3}\epsilon \,.
\end{equation}
For the odd-moment composite operator $O_s^{(2k-1)}$, the leading one-loop diagrams can be organized in the same way.
At the Nishimori fixed point, where the cubic couplings satisfy $u=3w$, these diagrams give the same anomalous contribution as for $O_s^{(2k)}$.
We therefore obtain
\begin{equation}
    \Delta_{2k-1}(d)=\Delta_{2k}(d),
\end{equation}
which is consistent with the Nishimori gauge-like symmetry discussed above.
This equality provides the perturbative counterpart of the multifractal pairing observed numerically in the main text.

The field theory also provides an understanding of the relevant perturbation away from the Dirichlet boundary condition.
Apparently, there are two relevant perturbations given by linear boundary term in $M$ or $Q$, 
\begin{eqnarray}
    \delta S_{\rm rel} = \int_{\partial \mathcal M} \left[ h \sum_\alpha M_\alpha - W \sum_{\alpha < \beta} Q_{\alpha\beta} \right] \,,
\end{eqnarray}
where $h$ corresponds to a non-random boundary field, whereas $W$ is from a random boundary field with zero mean. 
To derive the perturbation $W$, we consider a fixed boundary field $h(x)$. 
The replicated perturbation is $ \delta S_h =-\int_{\partial M}  h(x)\sum_\alpha \sigma_\alpha(x) $, where $\sigma_\alpha$ denotes the boundary spin field. 
Averaging over the zero-mean random field generates
\begin{eqnarray}
    \overline{e^{-\delta S_h}} = \exp\left[ \frac{W}{2} \int_{\partial \mathcal M} \left(\sum_\alpha \sigma_\alpha(x)\right)^2
    \right] \,,
\end{eqnarray}
where $W$ characterizes the variance. 
Expanding the square, the diagonal terms are constants for Ising spins, while the off-diagonal terms give the replica overlap field $Q_{\alpha\beta}\sim \sigma_\alpha\sigma_\beta$.  
Thus the leading nontrivial boundary perturbation is $- W \int_{\partial M} \sum_{\alpha<\beta} Q_{\alpha\beta}(x)$.
Therefore, the relevant exponents away from the free boundary fixed point are governed by the scaling dimension of the perturbation $y_h$ for the nonrandom field and $y_W$ for the random field:
\begin{eqnarray}
    y_h = d-1-\Delta_1 \,, \quad y_W= d-1-\Delta_2 \,.
\end{eqnarray}

This provides a nontrivial consistent check in our numerical calculation.  
In particular, we can derive from the microscopic boundary condition that
\begin{equation}
    \frac{e^{h}} {e^{-h}} = \frac{\cos(\pi/4-\theta/2)}{\sin(\pi/4-\theta/2)} \qquad \Rightarrow \qquad \tanh(h) = \tan(\theta/2) \qquad \Rightarrow \qquad h \approx \theta/2 \quad \text{if $\theta \ll 1$} \,.
\end{equation}
In this context, we can have a direct relation between the RG eigenvalue $y$ and the scaling dimension $\Delta_{1}$ or $\Delta_2$, which can be verified numerically as follows.

\begin{enumerate}
\item Uniform boundary perturbation: Microscopically, our boundary condition is parameterized by the spin-orientation angle $\theta$. 
For a small uniform angle $\theta \ll 1$, the boundary tensor $B_{\sigma}^{\theta}$ behaves as an effective magnetic field applied to the boundary spins. 
In the continuum limit, it deforms the replicated effective CFT action under the free boundary condition by an additional term $\sim h \int_{\partial \mathcal M} \sum_{\alpha} M_{\alpha}(x)$. 
Therefore, with $\Delta_{1} \approx 0.263$, we have the RG flow eigenvalue $y_h = 1 - \Delta_{1} \approx 0.737$, which is in good agreement with the value $y \approx 0.75$ obtained from the data collapse of the boundary entropy.

\item Random boundary perturbation: Similarly, we can apply this analysis to the binary random boundary perturbation. 
The perturbation introduces a random boundary magnetic field with zero mean and a variance $W \sim \theta_{W}^{2}$, leading to a deformation of the replicated effective action by $\sim W \int_{\partial \mathcal M} \sum_{\alpha \neq \beta}Q_{\alpha\beta}(x)$, with $\Delta_{2} \approx 0.263$.
This leads to the RG eigenvalue $y_{W} = 1 - \Delta_{2} \approx 0.737$\,. 
Finally, since the scaling variable is $\theta_{W} L^{y}$ instead of $\theta_{W}^{2} L^{y_{W}}$ in our data collapse, we should have $y = y_{W}/2 \approx 0.368$, which is also in good agreement with the value $y \approx 0.365$ obtained from the data collapse of the boundary entropy under the random boundary perturbation.

\end{enumerate}

\section{VI. Fermionic transfer matrix and coherent information}

\begin{figure}
    \centering
    \includegraphics[width=1.0\linewidth]{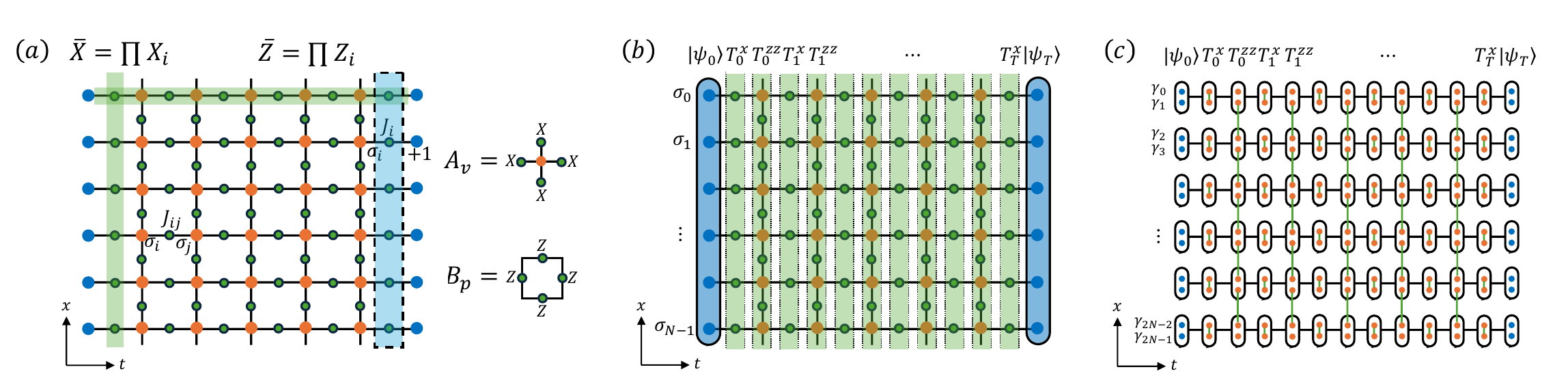}
    \caption{(a) Toric code and the corresponding RBIM representation for bit-flip errors.
    The toric code is defined with open boundary conditions along the $t$ direction and periodic boundary conditions along the $x$ direction.
    The green circles denote the qubits of the toric code, and the plaquette operators at the boundary are truncated to three-body operators.
    The two green lines represent logical operators, while the blue region marks the boundary qubits, which are assigned a distinct error rate $p_{\rm bdy}$.
    The orange dots denote the Ising spins $\sigma_i$ in the RBIM, and the green circles are mapped to the random couplings $J_{ij}$.
    The blue circles denote the boundary spins of the RBIM, which are fixed to $\sigma_i=1$ by the mapping from the toric code.
    (b) Spin representation of the fermionic transfer matrix.
    The blue regions denote the initial and final states of the transverse-field Ising model with random $ZZ$ couplings and random transverse-field $X$ terms.
    The green regions denote the elementary transfer matrices $T_t^x$ and $T_t^{zz}$.
    (c) Fermionic transfer matrix.
    After the Jordan-Wigner transformation from spins to Majorana fermions, the random $ZZ$ couplings are mapped to nearest-neighbor couplings between Majorana fermions, while the random transverse-field $X$ terms are mapped to two-Majorana terms on each site.}
    \label{fig:torci code to RBIM}
\end{figure}

In this section, we describe the fermionic transfer-matrix method used to evaluate the coherent information in the main text.
We consider the toric code on the geometry shown in Fig.~\ref{fig:torci code to RBIM}(a), with open boundary conditions along the $t$ direction and periodic boundary conditions along the $x$ direction.
With bit-flip noise $\mathcal{E}_i(\rho)=(1-p)\rho+pX_i\rho X_i$, a single-qubit error flips the plaquette operator $B_p=\prod_{i\in p} Z_i$ on the adjacent plaquettes.
In maximum-likelihood decoding, for a given error syndrome specifying which plaquette operators change sign, one sums over all physical error configurations that produce the same syndrome and leave the logical qubit invariant; this gives the probability $Z$.
Conversely, summing over all physical error configurations that produce the same syndrome but flip the logical qubit yields the probability $Z'$.
The decoder then compares $Z$ and $Z'$ and selects the more likely logical sector.

For example, when the syndrome contains two flipped plaquette operators, one may choose an arbitrary path connecting the two plaquettes without winding around the non-contractible cycle, and assign random couplings $J_{ij}=-J$ along this path.
All other physical error configurations with the same syndrome are then generated by flipping Ising spins on the vertices.
Therefore, each error syndrome defines an RBIM partition function with a fixed realization of random couplings.
The error-correction transition of the toric code with bit-flip noise is then mapped to the phase transition of the RBIM after summing over all possible random-coupling configurations.

The RBIM can further be mapped to a quantum $1+1$D transverse-field Ising model with random $ZZ$ couplings and random transverse-field $X$ terms, as shown in Fig.~\ref{fig:torci code to RBIM}(b).
In this mapping, the random $ZZ$ couplings correspond to the vertical bonds between spins, while the random transverse fields correspond to the horizontal bonds.
Each layer can then be represented by a transfer matrix $T_t^{zz}$ or $T_t^x$.
Thus the classical partition function can be written as a transfer-matrix evolution of a $1+1$D transverse-field Ising chain,
\begin{equation}
\label{eq:TFIM}
    Z\propto
    \left<\psi_T\right|
    T^x_T
    \prod_{t=0}^{T-1}
    T^{zz}_t T^x_t
    \left|\psi_0\right>,
\end{equation}
where
\begin{equation}
    T^x_t=
    \exp\left(\sum_i \widetilde J^t_{i,t}\sigma_i^x\right),
    \qquad
    T^{zz}_t=
    \exp\left(\sum_i J^x_{i,t}\sigma_i^z\sigma_{i+1}^z\right).
\end{equation}
Here $J^x_{i,t}$ denotes the coupling between sites $i$ and $i+1$ within the same time slice, while $\widetilde J^t_{i,t}$ denotes the coupling between adjacent time slices.
The dual coupling is $\widetilde J^t_{i,t}=\operatorname{arctanh}\left(e^{-2J^t_{i,t}}\right)$.
For binary disorder, $J^t_{i,t}=\eta^t_{i,t}J$ with $\eta^t_{i,t}=\pm1$, this can be written as $\widetilde J^t_{i,t}=\operatorname{arctanh}\left(e^{-2J}\right)+ {\rm i}\frac{\pi}{2}\delta_{\eta^t_{i,t},-1}$.

We next fermionize the transfer matrices using Majorana operators, as shown in Fig.~\ref{fig:torci code to RBIM}(c). 
With the convention
\begin{equation}
    \sigma_i^z=(-{\rm i})^i\prod_{j=0}^{2i}\gamma_j,
    \qquad
    \sigma_i^x=-{\rm i}\gamma_{2i}\gamma_{2i+1},
\end{equation}
one obtains $\sigma_i^z\sigma_{i+1}^z=-{\rm i}\gamma_{2i+1}\gamma_{2i+2}$ for $i<L_x-1$, and $\sigma_{L_x-1}^z\sigma_0^z={\rm i}\mathcal{P}\gamma_{2L_x-1}\gamma_0$ with parity $\mathcal{P}=\prod_i\sigma_i^x$.
After fixing the fermion-parity sector, both $T^x_t$ and $T^{zz}_t$ become Gaussian Majorana operators.
Here $T^{x}_t$ couples the two Majorana fermions on each site, while $T^{zz}_t$ couples Majorana fermions on nearest-neighbor sites.

It remains to specify the boundary states. 
For the coherent information of the toric code with the open boundary shown in Fig.~\ref{fig:torci code to RBIM}(b), the auxiliary boundary spins are fixed. 
For the untwisted partition function $Z$, we take the initial and final boundary configurations to be the same, 
\(\left|\psi_0\right>=\left|\sigma_i\equiv1\right>\) and 
\(\left|\psi_T\right>=\left|\sigma_i\equiv1\right>\). 
For the twisted partition function $Z'$, corresponding to the insertion of the logical operator $\overline X$, the signs of the boundary bonds are flipped, as shown in the blue region in Fig.~\ref{fig:torci code to RBIM}(a).
Equivalently, we keep the same disorder realization but flip the auxiliary spins on one boundary, so that 
\(\left|\psi_T\right>=\left|\sigma_i\equiv-1\right>\).

The fixed spin states do not have definite fermion parity, because $\mathcal{P}\left|\sigma_i\equiv1\right>=\left|\sigma_i\equiv-1\right>$.
We therefore introduce the parity-resolved combinations
\begin{equation}
    \left|\psi_{\rm fixed}^a\right>=\left|\sigma_i\equiv1\right>+a\left|\sigma_i\equiv-1\right>, \qquad a=\pm1 .
\end{equation}
Let \(Z^{a,b}\) denote the transfer-matrix amplitude with boundary states \(\left|\psi_{\rm fixed}^a\right>\) and \(\left|\psi_{\rm fixed}^b\right>\). 
The physical partition functions are then obtained as
\begin{equation}
    Z=\frac{Z^{1,1}+Z^{1,-1}}{4},
    \qquad
    Z'=\frac{Z^{1,1}-Z^{1,-1}}{4}.
\end{equation}
The states \(\left|\psi_{\rm fixed}^a\right>\) are Gaussian states: they can be represented as ground states of the quadratic Majorana Hamiltonian $H^a=-{\rm i}a\gamma_{2L_x-1}\gamma_0+{\rm i}\sum_{i=0}^{L_x-2}\gamma_{2i+1}\gamma_{2i+2}$, which imposes the stabilizer constraints generated by \(\{Z_iZ_{i+1},a\mathcal{P}\}\). 
Thus, both the transfer matrices and the boundary states admit a free-fermion representation.

Finally, the partition function can be evaluated as an overlap of Gaussian fermion states. 
Since the original classical partition function is positive, we avoid tracking possible Pfaffian signs by considering two identical copies and computing \(Z^2\). 
Introducing complex fermions $c_i=\frac{\gamma_i^{(1)}+{\rm i}\gamma_i^{(2)}}{2}$, the doubled Majorana bilinear becomes ${\rm i}\gamma_i^{(1)}\gamma_j^{(1)}+{\rm i}\gamma_i^{(2)}\gamma_j^{(2)}=2{\rm i}(c_i^\dagger c_j-c_j^\dagger c_i)$.
The Gaussian overlap is therefore reduced to a determinant,
\begin{equation}
\label{eq:Z two copy}
\begin{split}
    \left<\psi_T\right|T^x_T\prod_{t=0}^{T-1}T^{zz}_tT^x_t\left|\psi_0\right>=\sqrt{\left|\det\left(P_T^\dagger\mathbb{T}^x_T\prod_{t=0}^{T-1}\mathbb{T}^{zz}_t\mathbb{T}^x_tP_0\right)\right|} .
\end{split}
\end{equation}
Here \(P_0\) and \(P_T\) are the single-particle wavefunction matrices specifying the initial and final Slater determinants, and \(\mathbb{T}^{x,zz}\) are the corresponding single-particle transfer matrices. 
This determinant formula is the numerical scheme used to compute the partition functions entering the coherent information.

\begin{figure}
    \centering
    \includegraphics[width=0.4\linewidth]{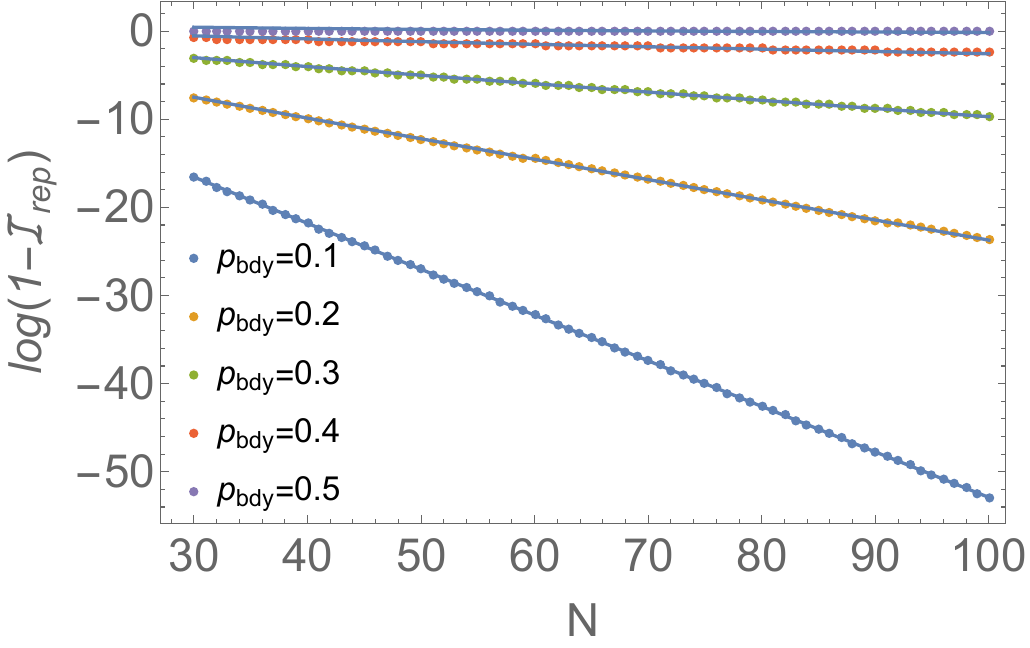}
    \caption{Coherent information of the boundary repetition code as a function of system size $N$. 
    Dots show results for boundary error rates $p_{\rm bdy}=0.1,0.2,0.3,0.4,0.5$, while blue lines are fits to Eq.~\eqref{eq:repetition code F CI app}. 
    For $p_{\rm bdy}<1/2$, $1-\mathcal{I}_{\rm rep}(p_{\rm bdy})$ decays exponentially with $N$, whereas $p_{\rm bdy}=1/2$ marks the boundary decoding transition.}
    \label{fig:CI numerics}
\end{figure}

In addition, we numerically compute the coherent information of the effective boundary code to verify the analytical result used in the End Matter of the main text. 
As shown in Fig.~\ref{fig:CI numerics}, we plot $\log(1-\mathcal{I}_{\rm rep})$ versus the system size $N$ for several boundary error rates $p_{\rm bdy}$. 
For all $p_{\rm bdy}<1/2$, the data decay exponentially with $N$. 
We fit the results using the asymptotic form
\begin{equation}
\label{eq:repetition code F CI app}
    \mathcal{I}_{\rm rep}(p_{\rm bdy})=1-A(p_{\rm bdy})N^{-1/2}\left[4p_{\rm bdy}(1-p_{\rm bdy})\right]^{N/2}+\cdots ,
\end{equation}
shown as blue lines in Fig.~\ref{fig:CI numerics}, and find excellent agreement with the numerical data.

\end{document}